\documentclass[%
 reprint,
nofootinbib,
 amsmath,amssymb,
 aps,
floatfix,
]{revtex4-2}

\usepackage{graphicx}% Include figure files
\usepackage{dcolumn}% Align table columns on decimal point
\usepackage{bm}% bold math
\usepackage{braket}
\usepackage{float}
\usepackage{subcaption}
\usepackage{underoverlap}
\usepackage{anyfontsize}
\usepackage{amsmath}
\usepackage{appendix}
\usepackage{xcolor}

\usepackage{silence}
\usepackage{ifthen}
\usepackage{tikz}
\usepackage{fourier}

\DeclareRobustCommand{\bigO}{%
  \text{\usefont{OMS}{cmsy}{m}{n}O}%
}
\DeclareMathOperator{\sgn}{sgn}

\begin{document}

\newcommand{\Qfactor}{Q} 
\newcommand{\Pfactor}{P} 

\preprint{APS/123-QED}

\title{Continuous Approximation of the Ising Hamiltonian:\\Exact Ground States and Applications to Fidelity Assessment in Ising Machines}%

\author{Amirhossein Rezaei$^1$}
%orcid id: {0000-0002-8937-8442}
\email{amirh.rezaei@alumni.sbu.ac.ir}

\author{Mahmood Hasani $^2$}%
\email{spew@aut.ac.ir}
\author{Alireza Rezaei $^2$}%
\email{alirezarezaei@aut.ac.ir}
\author{S.M. Hassan Halataei $^{1,}$}%
\email{m\_halataei@sbu.ac.ir}

\affiliation{%
$1$  Department of Physics, Shahid Beheshti  University, Evin, Tehran 19839, Iran}%

\affiliation{%
$2$
Department of
Electrical Engineering, Amirkabir University of Technology, Iran}%

\date{\today}% It is always \today, today,

\begin{abstract}

In this study, we present a novel analytical approach to solving large-scale Ising problems by reformulating the discrete Ising Hamiltonian into a continuous framework. This transformation enables us to derive exact solutions for a non-trivial class of fully connected Ising models. To validate our method, we conducted numerical experiments comparing our analytical solutions with those obtained from a quantum-inspired Ising algorithm and a quantum Ising machine. The results demonstrate that the quantum-inspired algorithm and brute-force method successfully align with our solutions, while the quantum Ising machine exhibits notable deviations. Our method offers promising avenues for analytically solving diverse Ising problem instances, while the class of Ising problems addressed here provides a robust framework for assessing the fidelity of Ising machines.  

\end{abstract}

% \keywords{Ising Model, Exact solution of Ground state Configuration and Energy, Coherent Ising Machine}%Use showkeys class option if keyword
\maketitle
%\tableofcontents
\section{Introduction} The Ising model \cite{brush1967history, barahona1982computational}, originated from statistical mechanics, is a mathematical model used to study and describe spin glasses. This model consists of a binary spin system with energy defined by the Ising Hamiltonian \cite{haribara2016computational}. A key problem of interest is determining the ground state energy of the Ising model. Specifically, the \textit{Ising problem} involves finding the configuration of $N$ Ising spins, $s_i = \pm 1$, that minimizes the Hamiltonian:
\begin{equation}\label{eq:fH}
H = \sum_{1 \le i < j \le N} J_{ij} s_i s_j,
\end{equation}
where a real number $J_{ij}$ denotes a coupling constant between every two of the $N$ Ising spins. Determining the ground state energy of the Ising model is, in general, an NP-hard problem \cite{knuth74}. This hardness depends on the specific choice of the couplings $J_{ij}$, not merely on the exponential number of configurations. The Ising model can encode any NP problem, and has been used to study NP-complete problems as well \cite{cipra2000}.
 This includes several classical problems such as the Max-Cut \cite{karp2010reducibility}, the Travelling Salesman Problem \cite{vcerny1985thermodynamical}, Set Cover \cite{karp2010reducibility}, Knapsack with Integer Weights \cite{salkin1975knapsack}, Graph Coloring \cite{jensen2011graph} and Clique Cover \cite{gramm2009data}.

Identifying and validating the exact ground state of the Ising Hamiltonian generally remains an unsolved problem. To find the ground state energy and configuration, Ising minimizers such as D-Wave quantum annealer \cite{king2023quantum, johnson2011quantum,king2024computational}, Coherent Ising Machine \cite{marandi2014network, inagaki2016large, mcmahon2016fully, inagaki2016coherent}, Bifurcation-based adiabatic quantum computation \cite{goto2016bifurcation, goto2019combinatorial, goto2021high} and Simulated Annealing \cite{kirkpatrick1983optimization,van1987simulated,bertsimas1993simulated} are used. However, a few specific analytical solutions also exist in the literature. \cite{huang2016finding} provides an exact provable solution for a  periodic lattice by transforming the problem into MAX-SAT and MAX-MIN optimization problems. \cite{dublenych2012ground} found the ground state for Shastry-Sutherland lattice in the presence of a magnetic field. Both of these solutions assume a finite interaction range, i.e., non-fully connected. Also, these solutions are only valid for uniform interaction couplings (where all elements have a fixed, equal value). Some notable works that go beyond these limitations are: \cite{bak1982one}, which demonstrated that the Ising model with long-range antiferromagnetic interactions exhibits a complete devil’s staircase. \cite{low1994study} also studied a system with competing short-range ferromagnetic coupling and long-range antiferromagnetic Coulomb interactions. In their study, they observed specific periodic configurations as the ground state.

Additionally, many studies of the Ising model rely on random matrix ensembles for benchmarking and theoretical analysis, such as Wishart-planted instances and other random coupling distributions \cite{perera2020chook, hamze2020wishart, hen2019equation}. However, real-world complex systems can often exhibit interaction strengths that are systematically determined by inherent properties or an underlying order of their components. In this paper, we explore such a scenario by introducing a class of fully connected Ising models where interaction strengths are a deterministic function of the spin indices themselves. We reformulate the Ising Hamiltonian as a continuous function, enabling a novel analytical approach to solving the models and determining their ground state.
To assess the effectiveness of this method, we conduct numerical experiments for validating our analytical solution, involving brute-force calculations, the Simulated Coherent Ising Machine \cite{ercsey2011optimization,yamamoto2017coherent,Chen_cim-optimizer_a_simulator_2022,zeng2024performance}(SimCIM) and the D-Wave quantum computer. This class of Ising Hamiltonian could serve as a framework for assessing the fidelity of Ising minimizers. The results demonstrate perfect agreement between brute-force calculations and our analytical approach, as well as between SimCIM and our method. However, the D-Wave quantum solver exhibits significant deviations for larger problem sizes.

To enhance the readability of the paper, we first present
our class of Ising models, followed by our method for solving them, and conclude with benchmarks and comparisons.

\section{The Interaction Matrix}    \label{sec:interaction}
  First, we introduce the interaction matrix:
\begin{equation}
\label{orderd_J}
    J_{ij}^{(N, d)}= \frac{1}{N^{d}}(i^d+j^d)(1-\delta_{ij}),
\end{equation}
where $i$ and $j$ are the indices of the matrix $J^{(N, d)}$, with $d$ being a real number. The variable $N$ is the size of the matrix. The terms $i^d$ and $j^d$ denote the indices $i$ and $j$ raised to the power of $d$, respectively. The interaction matrix in Eq.~(\ref{orderd_J}) thus implies an ordered and heterogeneous coupling structure. To offer an intuitive picture, one can liken this system to a city where individuals (indexed $1$ to $N$) possess an intrinsic, ordered 'economic power' or 'rank' related to their index $i$. The term $(i^d+j^d)$ then represents a combined 'influence potential' for an interaction between individuals $i$ and $j$. As we will demonstrate in subsequent sections, this structured heterogeneity, where interaction strengths are tied to the inherent ordering of the components, is key to the emergence of a predictable two-cluster ground state configuration, metaphorically, the 'city' organizing into two distinct, opposing 'poles'. As an example, for $N = 5$ and $d = 2$ we have:
\begin{equation}
    J^{(5, 2)} = \frac{1}{5^2}\begin{bmatrix}
    0 & 5 & 10 & 17 & 26 \\
    5 & 0 & 13 & 20 & 29 \\
    10 & 13 & 0 & 25 & 34 \\
    17 & 20 & 25 & 0 & 41 \\
    26 & 29 & 34 & 41 & 0 \\
    \end{bmatrix}.
\end{equation}

The ground state configuration for this class of interaction matrices is postulated to follow the pattern shown in Equation (\ref{eq:ground}). The proof substantiating this postulation will be provided in the subsequent section.

\begin{equation} \label{eq:ground}
    \pmb{s}_{g} = \left[ \begin{array}{cccccccc}
    1 \\ 1 \\ \vdots \\ 1 \\ -1 \\ -1 \\ \vdots \\ -1
    \end{array} \right].
\end{equation}

In this configuration, the up spins are adjacent to each other, as are the down spins. We denote the number of up spins by $M$, and the number of down spins by $N - M$. This means we can represent the ground state with only one variable, $M$. As an example, for the $J^{(5, 2)}$ the ground state is: $\pmb{s}_{g}^{(5, 2)} = \left[
    1, 1, 1, -1, -1
     \right],$
\noindent which has 3 up spins and 2 down spins. The Ising Hamiltonian (Equation (\ref{eq:fH})) is invariant under gauge transformation, i.e., all of the eigen-states of Equation (\ref{eq:fH}) are at least doubly-degenerate. In our notation, we consider the first cluster size as the up spin ($M$). 

\section{Ground State Pattern}\label{gsd1} 
 As $J^{(N, d)}$ is symmetric, the following is true:

\begin{equation} \label{eq:genhamil}
    H = \sum_{1 \le i < j \le N} J_{ij}^{(N, d)} s_i s_j = \frac{1}{2} \sum_{i=1}^N \sum_{j=1}^N J_{ij}^{(N, d)} s_i s_j.
\end{equation}

\begin{figure}[htbp!]
    \centering
    $\pmb{s_{g}}\pmb{s_{g}^T} =$
    $
    \begin{gathered}
    \includegraphics[width=0.55\columnwidth]{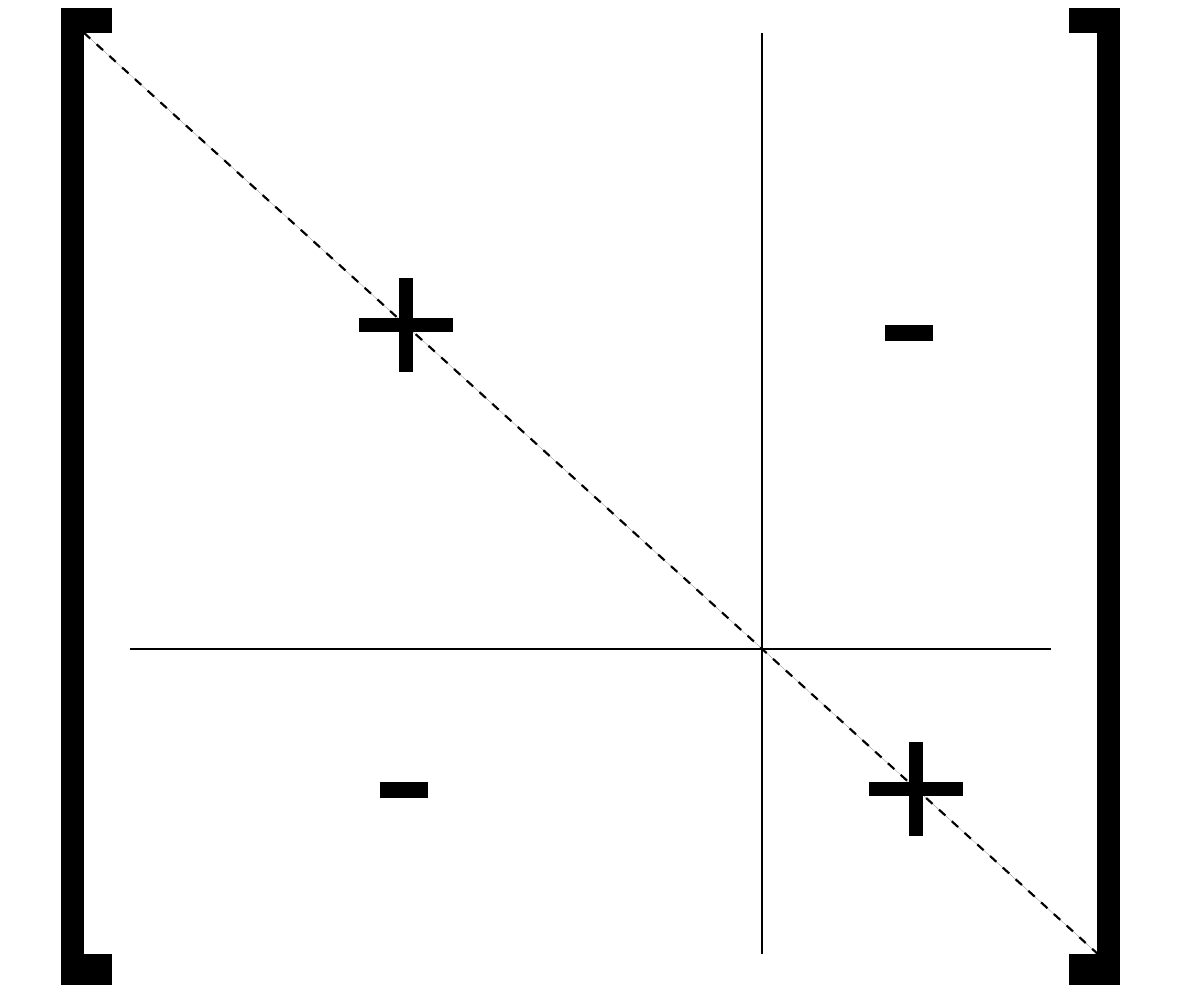} 
    \end{gathered}
    $
    \caption{Sign of interactions between spins, denoted by $\pmb{s}\pmb{s^T}$ and represented as a matrix.}
    \label{fig:interactions}
\end{figure}

Since the ground state configuration consists of $M$ adjacent up spins and $N-M$ adjacent down spins, the $\pmb{s}\pmb{s}^T$ matrix can be visualized as depicted in Figure \ref{fig:interactions}.

In this visualization, the upper left and lower right quadrants represent interactions between spins with the same orientation, i.e., $s_i s_j = 1$. Furthermore, the other two quadrants represent interactions between spins with different orientations, i.e., $s_i s_j = -1$.  Note that this diagram depicts the sign of the interactions between spins, not the coupling values themselves. 

If we assume the ground state is represented by Equation (\ref{eq:ground}), we can proceed as follows. By considering the upper left quadrant as an $M \times M$ matrix, we can rewrite the Hamiltonian (Equation (\ref{eq:genhamil})) in the following form:
    \begin{multline}
         \label{eq:hamil}
 H(M, N, d) = \frac{1}{2N^{d}}\{\sum_{1 \le i \neq j \le M} (i^d+j^d) + \sum_{M+1 \le i \neq j \le N} (i^d+j^d)  - \\
         \sum_{{i=1}}^M \sum_{{j=M+1}}^N (i^d+j^d) - \sum_{{i=M+1}}^N \sum_{{j=1}}^M (i^d+j^d)\},
    \end{multline}

\noindent where the first and second terms correspond to interactions between spins which have the same orientation, i.e., $s_is_j = 1$. The third and fourth terms correspond to the interactions between spins with opposite orientation such that: $s_is_j = -1$.

Considering the following formula \cite{knuth1993johann}:

\begin{equation}\label{faulhaber}
    F^{d}(N) = \sum_{{i=1}}^N i^d = \sum_{r=0}^{d} \frac{(-1)^r B_r}{d+1} \binom{d+1}{r} N^{d+1-r},
\end{equation}
\noindent where $B_r$ is the $r$th Bernouli number. We can write Equation (\ref{eq:hamil}), in terms of Equation (\ref{faulhaber}) as follows:
\begin{equation} \label{eq:hamilhaber}
    H(M,N,d) = \frac{1}{N^{d}}\left((N - 2M - 1)F^d(N) + (4M-2N)F^{d}(M)\right).
\end{equation}

Now to find the ground state energy and configuration, we just need to find the $M$ that minimizes $H$:
    \begin{equation}\label{argminM}
        M_{g}^{(N, d)} = \operatorname*{argmin}_M H(M, N, d),
    \end{equation}

\noindent Given that $M$ is an integer within the range $[1, N]$, the complexity of the problem becomes $\bigO(N)$, which is obviously polynomial. 

As $N$ grows, the ratio $\frac{M}{N}$ stabilizes to a constant. We denote this constant ratio as $q$. As $q$ varies by unit of $\frac{1}{N}$, in the limit of large $N$ this change becomes very small, thus we can treat $q$ as a continuous variable. Our next step is to determine the value of $q$ in the large $N$ limit. To do so, we minimize the function $H$ with respect to $q$:

\begin{equation}\label{derivative}
    \frac{\partial H(M, N, d)}{\partial q} = 0.
\end{equation}

In Equation (\ref{eq:hamilhaber}), by considering terms which are dependent on M and discarding $\frac{-2}{N^{d}}$, we only need to minimize Equation (\ref{htildefauhlber}) as written below:

\begin{equation}\label{htildefauhlber}
    \tilde{H}(M, N, d) = M F^{d}(N) + (N-2M)F^{d}(M).
\end{equation}

Using $M = qN$, Equations (\ref{derivative}) and (\ref{htildefauhlber}) we arrive at:

\begin{equation} \label{equality}
    N \{F^{d}(N) - 2 F^{d}(M) + (1-2q) \frac{\partial F^{d}(M)}{\partial M}\frac{\partial M}{\partial q}\} = 0.
\end{equation}

Derivative of Equation (\ref{faulhaber}) is:
\begin{equation} \label{df}
    \frac{\partial F^{d}(M)}{\partial M} = d F^{d-1}(M) + (-1)^{d} B_d.
\end{equation}

Substituting Equation (\ref{df}) in Equation (\ref{equality}) we get:

\begin{equation}\label{par}
    F^{d}(N) - 2 F^{d}(M) + (1-2q)N \{dF^{d-1}(M) + (-1)^{d} B_d \}  = 0.
\end{equation}

Keeping the the first leading order term in Equation (\ref{faulhaber}) results in:
\begin{equation}
    F^{d}(N) \approx \frac{N^{d+1}}{d+1}B_0,
\end{equation}

\noindent and substituting it in Equation (\ref{par}), we can write as below:
\begin{equation}
     \frac{B_0N^{d+1}}{d+1}  (1-2 q^{d+1}) + (1-2q)N((qN)^dB_0+(-1)^dB_d)=0.
\end{equation}

If we multiply both sides by $\frac{d+1}{B_0 N^{d+1}}$, for large $N$, $\frac{B_d}{N^{d+1}}$ tends to zero and we arrive at the final equation:
\begin{equation} \label{eq:12}
    1 + (1+d) q^d - 2 (2+d) q^{d+1} = 0.
\end{equation}

Solving this equation, gives us the value of $q$ which fully determines the ground state of the Ising model at large $N$. For $d \notin \{1, 2, 3\}$, Equation (\ref{eq:12}) becomes a transcendental equation and does not have a closed form solution. In these cases, we resort to numerical methods (such as the Newton-Raphson method) to find the roots of the equation.
\section{Theoretical Approach to Find the Ground State Pattern} 
Now note that \textit{any} spin configuration could be represented as Equation (\ref{eq:config}), which consists of an arbitrary number of domains of up and down spins. Each domain can contain one or more spins of the same orientation:
\begin{equation} \label{eq:config}
    \pmb{s}^T = [ \UOLunderbrace{1 \; 1 \cdots 1 \;}_{c_1} \;
    \UOLunderbrace{\text{-}1 \; \text{-}1 \cdots \text{-}1\;}_{c_2}\;
    \cdots \;
    \UOLunderbrace{1 \; 1 \cdots 1 \;}_{c_{\Lambda}} \;
    \UOLunderbrace{\text{-}1 \; \text{-}1 \cdots \text{-}1\;}_{c_{\Lambda+1}}].
\end{equation}

In continuous limit, the following function is equivalent to Eq. (\ref{eq:config}):
\begin{equation} \label{s(x)}
    S(x, \mathbf{q}) = (-1)^\Lambda \prod_{\alpha=1}^{\Lambda} \sgn(x-q_{\alpha}).
\end{equation}

Figure \ref{fig:config}, is a visualization of Equation (\ref{s(x)}). Each spin domain, denoted by $c_i$ with $i \in [1,\Lambda+1]$, occupies a certain region whose right boundary is denoted by $q_{\alpha}$ with $\alpha \in [1,\Lambda]$:
\begin{figure}

    \centering
    \definecolor{navy}{RGB}{0,0,128}
    \resizebox{1.\columnwidth}{!}{
    \begin{tikzpicture}
    \node[left] at (0,1) {1}; % Label the point 1 on the y-axis
    \node[left] at (0,0) {$q_0=0$}; % Label the point 0 on the x-axis
    \node[left] at (0,-1) {-1}; % Label the point -1 on the y-axis
    \draw (-0.1,1) -- (0.1,1); % Draw tick mark for 1 on the y-axis
    \draw (-0.1,-1) -- (0.1,-1); % Draw tick mark for -1 on the y-axis
    \draw[thick,->] (0,-2) -- (0,2); % Draw the y-axis
    \draw[thick,-] (0,0) -- (3.25,0); % Draw the x-axis
    \draw[thick,->] (4.25,0) -- (10,0); % Draw the x-axis
    \node at (3.75,0) {$\cdots$}; % Add ellipsis in the middle
    \node at (3.75,1) {$\cdots$};
    \node at (3.75,-1) {$\cdots$};
    % Add ellipsis in the middle
    \node[left] at (0,2) {$S(x, \mathbf{q})$};
    \node[below right] at (10,0.2) {$x$};
    \def\prevx{0}
    \foreach \x/\y/\z [remember=\x as \prevx] in {0.7/c_1/q_1, 2.5/c_2/q_2, 5/c_{k-2}/q_{\Lambda-3}, 6.3/c_{k-1}/q_{\Lambda-2}, 7.1/c_{k}/q_{\Lambda-1}, 8/c_{k+1}/q_{\Lambda}, 9/q_{\Lambda+1}=1} % Loop over each section and point
    {
        \node[below] at (\x,-0.12) {\scriptsize $\z$}; % Label the points
        \draw (\x,-0.1) -- (\x,0.1);
    }

\tikzset{mydashed/.style={navy, thin, dashed, dash pattern=on 1pt off 1pt, opacity=0.4}}

% First step (solid horizontal and dashed vertical lines)
\draw[navy, ultra thick] (0,1) -- (0.7,1); % Solid horizontal
\draw[mydashed] (0.7,1) -- (0.7,-1); % Dashed vertical
\draw[navy, ultra thick] (0.7,-1) -- (2.5,-1); % Solid horizontal
\draw[mydashed] (2.5,-1) -- (2.5,1); % Dashed vertical
\draw[navy, ultra thick] (2.5,1) -- (3.25,1); % Solid horizontal

% Second step (solid horizontal and dashed vertical lines)
\draw[navy, ultra thick] (4.25,1) -- (5,1); % Solid horizontal
\draw[mydashed] (5,1) -- (5,-1); % Dashed vertical
\draw[navy, ultra thick] (5,-1) -- (6.3,-1); % Solid horizontal
\draw[mydashed] (6.3,-1) -- (6.3,1); % Dashed vertical
\draw[navy, ultra thick] (6.3,1) -- (7.1,1); % Solid horizontal
\draw[mydashed] (7.1,1) -- (7.1,-1); % Dashed vertical
\draw[navy, ultra thick] (7.1,-1) -- (8,-1); % Solid horizontal
\draw[mydashed] (8,-1) -- (8,1); % Dashed vertical
\draw[navy, ultra thick] (8,1) -- (9,1); % Solid horizontal
\draw[mydashed] (9,1) -- (9,-1); % Dashed vertical

\end{tikzpicture}}

    \caption{Visualization of arbitrary spin configurations in continuous form. $q_\Lambda$ denote the boundaries of each cluster, which consists of spins with the same orientation. In this figure $q_0 = 0$ and $q_{\Lambda + 1} = 1$.}
    \label{fig:config}
\end{figure}
the set of boundaries $q_{\alpha}$ can also be represented as a vector and  for convenience, let's put $q_0 = 0$ and $q_{\Lambda+1} = 1$ which can be denoted by Equation (\ref{eq:ratio})
\begin{equation}\label{eq:ratio}
    \mathbf{q} = (q_1,q_2,..,q_{\Lambda}) \quad\text{,}\quad 0 < q_{\alpha} < 1 \quad\text{,}\quad q_{\alpha+1} > q_{\alpha}.
\end{equation}

Now, by neglecting the Kronecker delta in Equation (\ref{orderd_J}) which just adds a constant term to the Hamiltonian, we can write as below:
\begin{equation}\label{tilde}
     H = \frac{1}{2} \sum_{i=1}^N \sum_{j=1}^N J_{ij}^{(N, d)} s_i s_j = \frac{N^2}{2}\sum_{i=1}^N \sum_{j=1}^N\left((\frac{i}{N})^d+(\frac{j}{N})^d\right) s_i s_j\Delta i \Delta j,
\end{equation}
where $\Delta i = \Delta j = \frac{1}{N}$.
Summing Equation (\ref{tilde}) over indices $i$ and $j$ is equivalent to a Riemann sum, which is defined as:
% \lim_{max ||\Delta x_i||\to 0}
\begin{equation} \label{eq:reimann}
     \sum_{i=1}^N f(x_i^*) \; \Delta x_i = \int_{a}^{b} f(x) \; dx + \bigO \left(\frac{(f(b)-f(a))(b-a)}{N}\right),
\end{equation}
where $\Delta x_i = x_i - x_{i-1}$ and $x_i^* \in [x_{i-1}, x_i]$ and $x_0 = a< x_1 < x_2 < ... < x_{N-1} < x_N = b$. Thus, it can be treated as an integral by setting $(\frac{i}{N}) = x$ and $(\frac{j}{N}) = y$ and $\Delta x = \frac{1}{N}$  and taking its limit for large $N$ with an error of order $\bigO(\frac{1}{N})$. We should note that for smooth functions, this method works appropriately but for not-integrable functions the error term in Equation (\ref{eq:reimann}) diverges. By replacing $s_i$ and $s_j$ in Equation (\ref{tilde}) with $S(x,\mathbf{q})$ and $S(y,\mathbf{q})$ respectively, which are defined in Equation (\ref{s(x)}) and using Equation (\ref{eq:reimann}) twice:
\begin{multline} \label{eq:hintegral}
    H_{\Lambda}(d, \mathbf{q})  = \frac{N^2}{2}\times \\
    \int_{0}^{1}\int_{0}^{1} \{ (x^d + y^d) \prod_{\alpha=1}^{\Lambda} \sgn(x-q_{\alpha}) \sgn(y-q_{\alpha}) \} \; dx \; dy,
\end{multline}
where the term $\prod_{\alpha=1}^{\Lambda} \sgn(x-q_{\alpha}) \sgn(y-q_{\alpha})$ is the continuous form of $\pmb{s}\pmb{s}^T$ and $y_i = \frac{i}{N}$. Moreover, using Equation (\ref{eq:reimann}) to obtain Equation (\ref{eq:hintegral}) we have an error term which can be neglected in this problem due to large $N$, thus we can not use this method for any arbitrary Ising problem for finding the ground state pattern if the derivative of the integrand does not exist. Note that this integral only has solution for $d>-1$. Evaluation of Equation (\ref{eq:hintegral}) leads to Equation (\ref{eq:finalhamil}), where the details can be followed in Appendix \ref{Evaluating Integral}:
\begin{multline}\label{eq:finalhamil}
    H_{\Lambda}(d, \mathbf{q})  = \frac{N^2}{1 + d}\times\\
    \left((-1)^{\Lambda} + 2\sum_{\alpha=1}^{\Lambda} (-1)^{\alpha+1}q_\alpha\right)\left((-1)^{\Lambda} + 2\sum_{\alpha=1}^{\Lambda} (-1)^{\alpha+1}q_\alpha^{d+1}\right).
\end{multline}

The structure of $H_{\Lambda}$ in Eq.~(\ref{eq:finalhamil}) can be written as $H_{\Lambda} = \frac{N^2}{1+d} \Qfactor \Pfactor$, where $\Qfactor = ((-1)^{\Lambda} + 2\sum_{\alpha=1}^{\Lambda} (-1)^{\alpha+1}q_\alpha)$ and $\Pfactor = ((-1)^{\Lambda} + 2\sum_{\alpha=1}^{\Lambda} (-1)^{\alpha+1}q_\alpha^{d+1})$. This form is key to understanding the system's energetic preference for a ground state consisting of only two spin clusters (i.e., $\Lambda=1$). The energy is minimized when the product $\Qfactor \Pfactor$ is maximally negative.

For the $\Lambda=1$ case, the Hamiltonian simplifies to $H_1 = \frac{N^2}{1+d} (2q_1-1)(2q_1^{d+1}-1)$. As derived in Appendix~\ref{app:unique}, by substituting the condition for the critical value of $q_1$ (from Eq.~(\ref{eq:h1derive})) back into this expression, we find $H_1(d,q_1) = -N^2(2q_1-1)^2q_1^d$. This explicitly demonstrates that $H_1$ is always negative for $q_1 \neq \frac{1}{2}$ and $d>-1$, achieving a well-defined minimum due to the convexity of $H_1$ in the relevant intervals of $q_1$ and the uniqueness of the critical $q_1$ (see Appendix~\ref{app:unique} for details). Conversely, for configurations with multiple domain walls ($\Lambda \ge 2$), Appendix~\ref{app:pattern} details two crucial aspects. Firstly, if such a configuration possesses an interior critical point (where $\partial H_\Lambda / \partial q_j = 0$ for all $j$), it must satisfy the condition $\Qfactor=0$, as shown by Eq.~(\ref{eq:condition}) in Appendix~\ref{app:pattern}. This directly forces the total energy $H_\Lambda = 0$. An energy of zero is clearly less favorable than the negative energy achievable by the $\Lambda=1$ state. Secondly, any configuration with $\Lambda \ge 2$ domain walls can reduce its number of interfaces by allowing a domain boundary parameter $q_j$ to approach its limit (e.g., $q_\Lambda \to 1$). In such cases, the Hamiltonian $H_\Lambda$ mathematically transforms into $H_{\Lambda-1}$, effectively merging two domains. This process, detailed in Appendix~\ref{app:pattern}, can be iterated until the $\Lambda=1$ state is reached.

Therefore, the energetic preference for $\Lambda=1$ arises because it is the only configuration that allows the collective factors $\Qfactor$ and $\Pfactor$ to cooperate optimally to produce a significantly negative energy. Additional domain walls ($\Lambda \ge 2$) either disrupt this cooperation by forcing $\Qfactor=0$ (leading to $H_\Lambda=0$) or represent unstable, higher-energy states that can relax by shedding interfaces until the robustly negative energy of the $\Lambda=1$ state is attained. This ensures that the ground state pattern consists of just two clusters. The full mathematical derivations supporting these conclusions are presented in Appendices~\ref{app:unique} and \ref{app:pattern}.

\begin{figure}[hbtp!]
\centering
\includegraphics[width=\columnwidth]{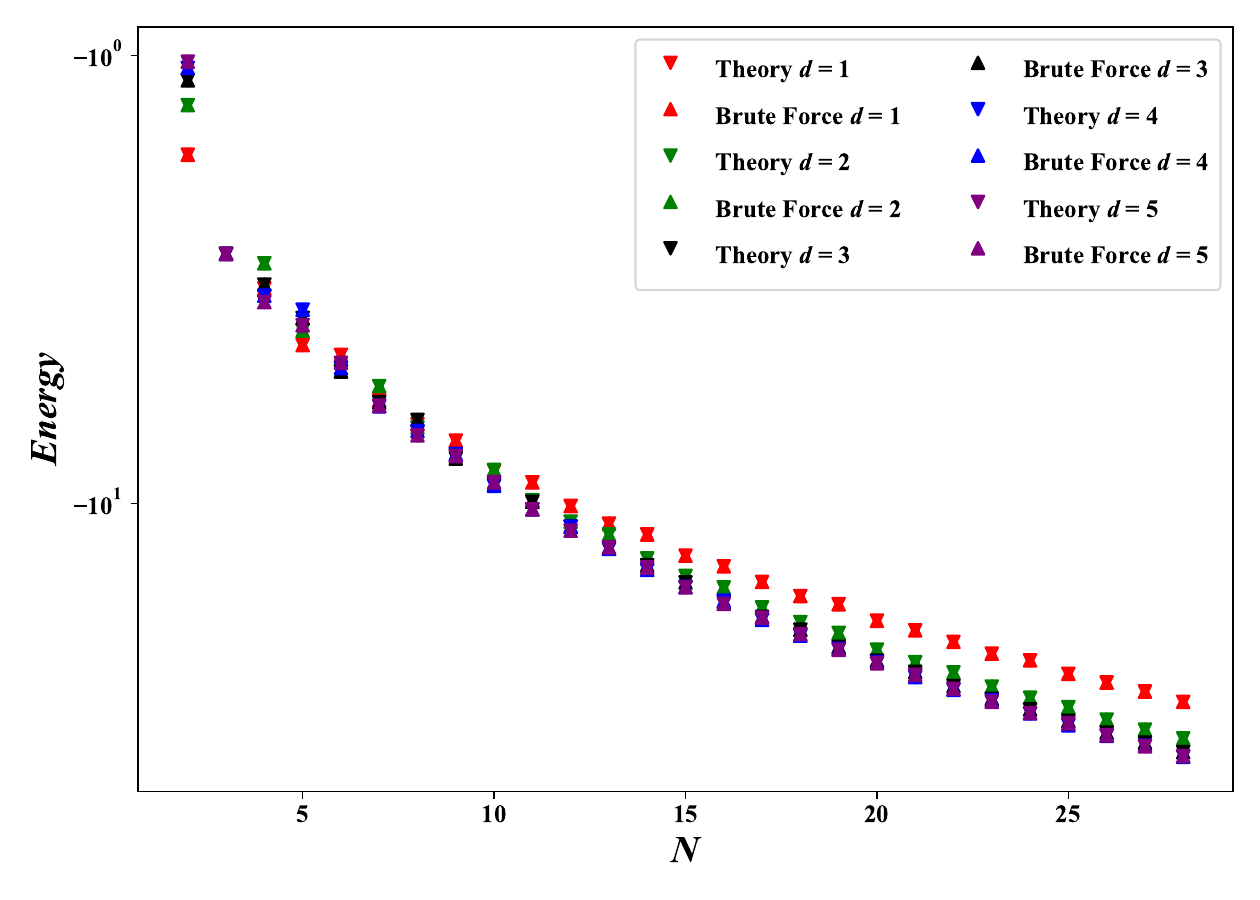}
  \caption{The ground state energy, computed for $d\in[1,5]$ and $N\in[1,28]$, using brute force approach (up triangle) and by obtaining the value of $M$ from Equation (\ref{argminM}) and substituting in Equation (\ref{eq:hamilhaber}) (down triangle).}
  \label{fig1}
\end{figure}

\begin{figure}[hbtp!]
\centering
\includegraphics[width=\columnwidth]{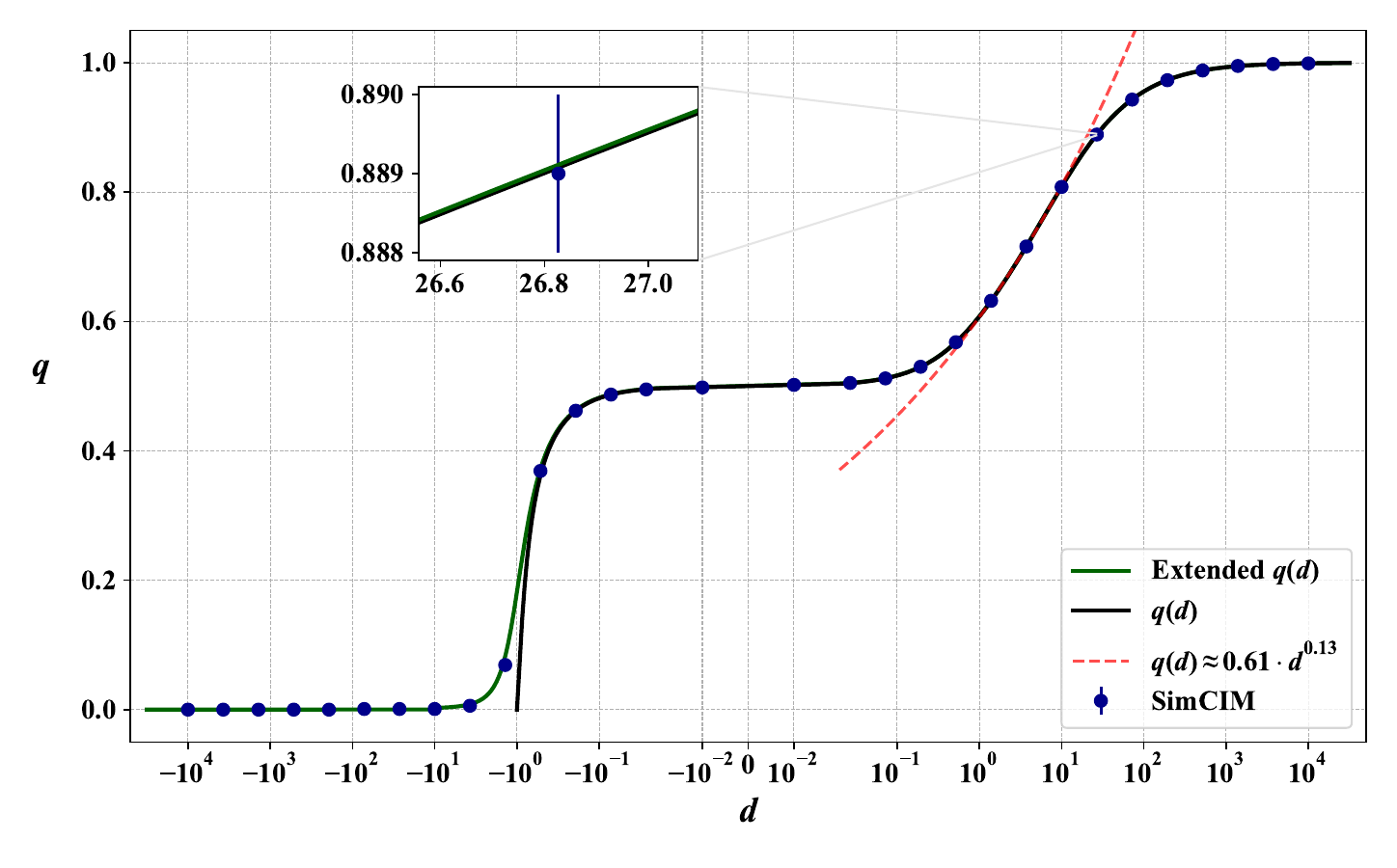}
  \caption{
  In the ground state configuration, the up spins are adjacent to each other, as are the down spins. We denote the number of up spins as $M$. This means we can represent the ground state with only one variable, $M$. For large $N$, the ratio $q$ stabilizes to a constant. This figure, depicts the plot of $q$ against $d$ ($-10^{4}$ to $10^{4}$). Dark line: root of Equation (\ref{eq:12}). Green line: root of Equation (\ref{extended_q}). Blue dots: SimCIM results. Error bars denote $\frac{1}{N}$ precision ($N = 1000$). Power law is observed for $d \in [1, 10]$, followed by saturation towards 1. At $d = 10$, $q \approx 0.81.$}
\label{fig2}
\end{figure}

\section{Numerical Results}
\subsection{Simulated Coherent Ising Machine and Brute-Force Method} \label{subsec:cim}
Now we calculate numerical results regarding the ground state of the system, described by Equation (\ref{eq:hamilhaber}). In Figure \ref{fig1}, we employed theoretical calculations for \( d = 1, 2, 3, 4, 5 \). To validate our results, we also conducted brute force search to determine the ground state of the system described by Equation (\ref{eq:hamilhaber}) for each \( d \) value, across Ising problem sizes ranging from 2 to 28. Figure \ref{fig1} shows that the energy values of our calculations and brute force search are exactly aligned. For larger systems, determining the ground state via brute force becomes infeasible due to the exponential increase in computational resources required.

For larger Ising problem sizes, particularly for $N = 1000$ as depicted in Figure \ref{fig2}, we exploited a Simulated Coherent Ising Machine. The results from this simulation, which completely agree with Equation (\ref{eq:12}), were obtained using the Chaotic Amplitude Control (CAC) algorithm. Details of the hyperparameter tuning of CAC is presented in the Appendix \ref{appendixa}.

Figure \ref{fig2} depicts the values of $q$ for $d$ ranging from $-10^{4}$ to $10^{4}$. The dark and green line represent the root of Equation (\ref{eq:12}) and Equation (\ref{extended_q}) respectively, while the blue dots correspond to the results obtained through the SimCIM. Given that our simulations were conducted with $N = 1000$, the precision of the $q$ ratio is limited to three decimal places. Consequently, the blue error bars are set to $\frac{1}{1000}$, reflecting the precision of $\frac{1}{N}$ for any given $N$. It is observed that $q(d)$ starts at $d = -1$ and then rapidly grows and stays near $\frac{1}{2}$ for $ d\in [-0.1, 0.1]$, then follows a power law for $d \in [1,10]$, and then saturates and tends to 1. We fitted a line for $d \in [1,10]$ (on a log-log scale), and it follows $q(d) = 0.61 \cdot d^{0.13}$. As the interaction matrix is ordered and follows a hierarchical pattern, this likely underlies the observed power law trend for $d$ values greater than 1 and less than 10. At $d = 10$, where the saturation begins and the power law behavior ends, the value of $q$ is approximately 0.81.
\subsection{D-wave Benchmarking and Fidelity Analysis} \label{sec:dwave}
\begin{figure}[ht!]
    \centering
    \includegraphics[width=1.\linewidth]{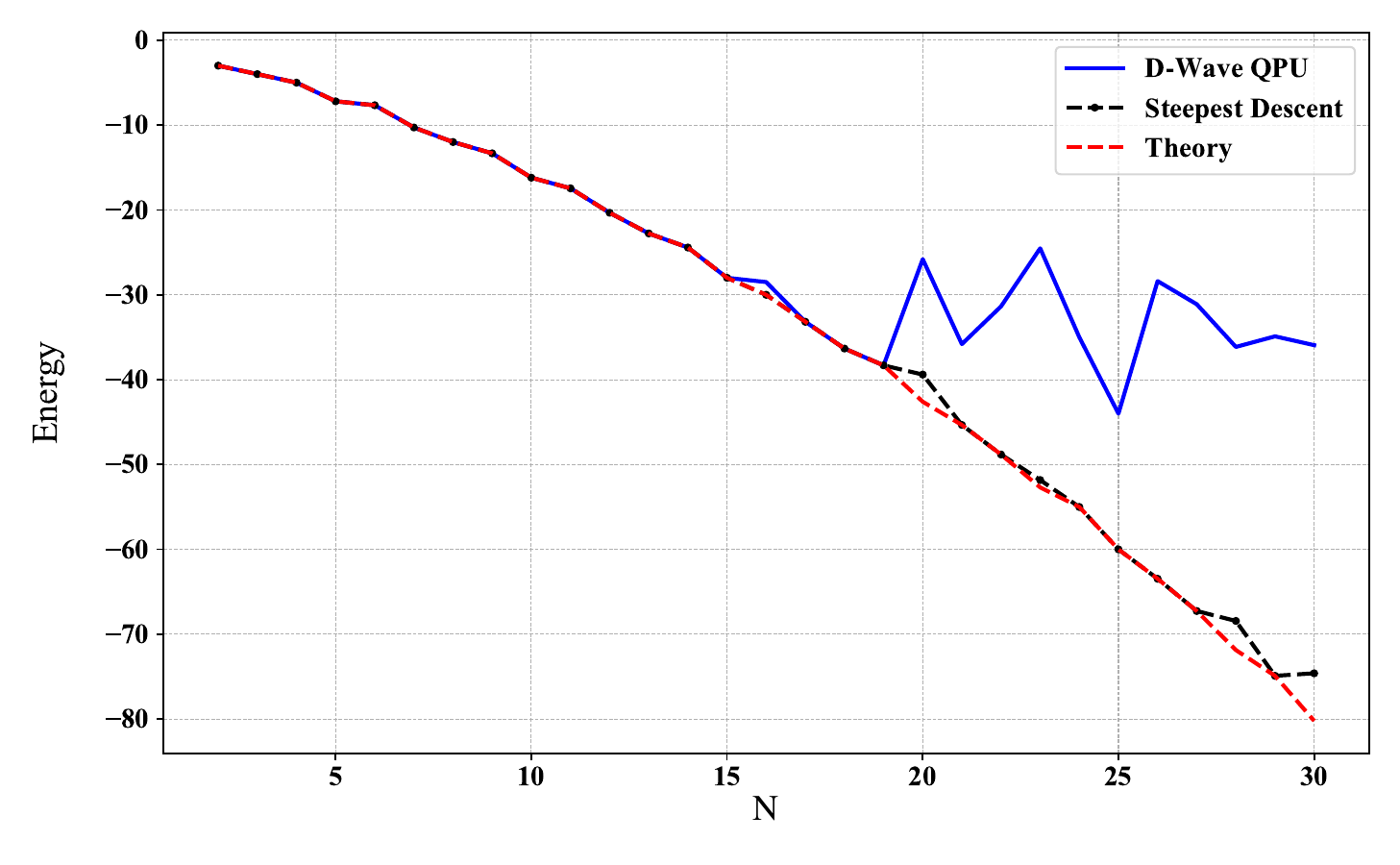}
    \caption{Benchmarking results of the D-Wave quantum annealer and the steepest descent algorithm for problem sizes in the range $[2, 30]$ with $d = 1$. For each problem size on D-Wave QPU, we obtained 1500 samples, with an annealing time of 1~$\mu s$. The energy values are compared against theoretical predictions. The D-Wave QPU shows significant deviation from theoretical values starting at $N = 20$. The deviations shown here for $d=1$ are indicative of the D-Wave QPU's performance for this class of problems. Similar significant deviations from theoretical values have been observed in our analyses for other $d$ values, including $d < 0$. This observation is consistent with the qualitative quasi-symmetry of the $J^{(N,d)}$ energy landscape around $d=0$ (Figure \ref{fig:full_spectrum_J_nd_panel} and supplementary video\cite{supplementary}), which means landscapes for $d<0$ present similar structural complexities and can be equally or more rugged (Figure \ref{fig:local_minima}). Since the QPU's performance limitations for these fully-connected problems are primarily due to encoding fidelity challenges (such as chain breaks, discussed at the end of Section \ref{sec:dwave}), these difficulties persist across different $d$ values. While the problem is polynomial, the D-Wave QPU cannot reach the ground state, making it a useful benchmark for assessing the fidelity of the QPU. The steepest descent algorithm generally reaches the ground state but sometimes gets trapped in local minima very close to the true ground state.}
    
    \label{fig:dwave}
\end{figure}

The performance of Ising solvers, such as the D-Wave quantum annealer, depends not only on their ability to optimize complex energy landscapes but also on how accurately the Ising Hamiltonian is encoded into the physical hardware itself. Errors in encoding, noise, or hardware imperfections can significantly degrade solution quality. To assess this aspect independently, we employed our benchmark interaction matrix \( J^{(N, d)} \), which has an analytical ground state. This matrix serves as an ideal tool to evaluate the fidelity of encoding while isolating it from the solver’s optimization capabilities.  

By leveraging \( J^{(N, d)} \), we were able to directly compare the results from the D-Wave system with exact theoretical solutions. Specifically, we used the \textit{D-Wave Advantage system}\cite{mcgeoch2020d}, which features over 5,000 superconducting qubits connected via the \textit{Pegasus} topology. This setup enabled us to assess how accurately the D-Wave hardware maps the mathematical problem onto its physical qubits, providing insights into potential limitations stemming from encoding errors. This method allows for a clearer distinction between encoding fidelity and the solver’s capacity to minimize energy.

Figure~\ref{fig:dwave} presents the benchmarking results for the D-Wave system alongside the Steepest Descent (SD) algorithm across problem sizes ranging from \( N = 2 \) to \( N = 30 \), with \( d = 1 \). For smaller problem sizes (\( N \leq 20 \)), the D-Wave results align closely with both SD and the exact theoretical solutions. However, as the problem size increases (\( N > 20 \)), deviations become apparent, highlighting the growing difficulty of maintaining encoding accuracy as the complexity of the fully connected problem increases. These discrepancies underscore the inherent challenges of mapping complex problems onto D-Wave hardware with perfect fidelity. Contributing factors include precision loss due to the finite quantization step of the digital-to-analog converter (DAC), unintended couplings arising from background susceptibility, and pervasive noise sources such as flux noise. A particularly significant issue for \( N > 20 \) is the increased incidence of \emph{chain breaks}, where noise disrupts the qubit chains required for problem embedding, resulting in inconsistent readouts and degraded solution quality.

In contrast, as discussed in Section (\ref{subsec:cim}), the Simulated Coherent Ising Machine (SimCIM) demonstrated the capability to consistently reach the ground state for problem sizes as large as \( N = 1000 \). As a simulated algorithm, SimCIM operates without the hardware-based encoding limitations inherent in physical systems like D-Wave. This emphasizes the crucial role of encoding fidelity in benchmarking physical quantum annealers, as SimCIM’s performance was unaffected by such constraints.  

The SD algorithm, often used as a post-processing tool for D-Wave results, was also evaluated as a standalone solver. Its solutions closely track the theoretical ground states across all problem sizes, indicating that the deviations observed in D-Wave results stem primarily from encoding fidelity rather than optimization performance.

\section{Energy Landscape Analysis}
\label{sec:energy_landscape}

To better understand the challenges that the Ising Hamiltonian 
Equation (\ref{eq:fH}) with the interaction matrix Equation (\ref{orderd_J}) poses to heuristic solvers and physical annealers, we now turn to an analysis of its energy landscape. While our continuous approximation method efficiently determines the global ground state configuration and energy, the characteristics of the broader landscape, including the number and distribution of local minima, are not directly revealed by this approach. These features are critical, as they heavily influence the performance of solvers that navigate the solution space based on local energy gradients or quantum tunneling, and a complex landscape with numerous local optima is a hallmark of computationally hard problems.

\subsection{Number of Local Minima}
\label{subsec:num_local_minima}

To identify local minima in the energy landscape defined by the coupling matrix $J^{(N,d)}$, we exhaustively enumerate all $2^N$ spin configurations and compute their energies. Each configuration is then compared to its $N$ single–spin–flip (Hamming‐distance‑1) neighbors: if its energy is lower than that of all its neighbors, it is classified as a local minimum. This exact procedure ensures a complete count of all local minima in the system for sizes up to certain problem size.

\begin{figure}[htbp!]
    \centering
    \includegraphics[width=\columnwidth]{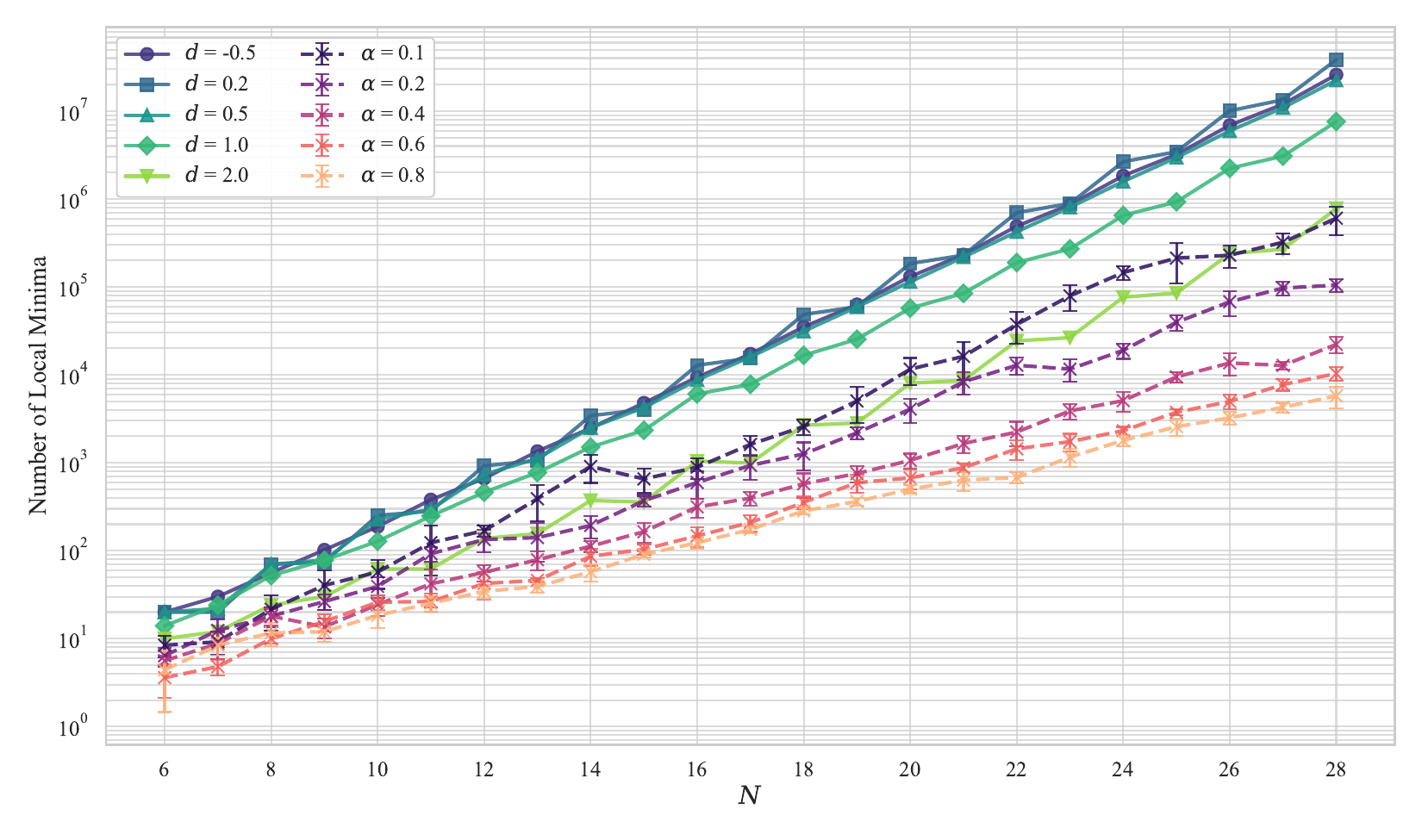} % Replace with your image file
    \caption{Number of local minima as a function of system size $N$. Solid lines show results for our proposed interaction matrix $J^{(N,d)}$ Equation (\ref{orderd_J}) for different values of $d$, as indicated in the legend. Dashed lines show corresponding data for the Wishart planted ensemble (parameterized by $\alpha$) from \cite{hamze2020wishart}. Local minima were found by exhaustively checking all $2^N$ configurations for stability against single spin flips. Error bars for Wishart data indicate the standard error over 100 instances. Notably, $J^{(N,d)}$ with smaller $d$ values can exhibit a larger number of local minima than the Wishart instances for similar $N$.}
    \label{fig:local_minima}
\end{figure}

Figure~\ref{fig:local_minima} shows the number of local minima found using this brute-force approach for $J^{(N,d)}$ with varying $d$ and for the Wishart planted ensemble with varying $\alpha$.
A prominent feature for all cases is the apparent exponential growth in the number of local minima with increasing system size $N$. Such exponential proliferation of metastable states is a strong indicator of a rugged energy landscape and is characteristic of many computationally hard optimization problems, including NP-hard problems like spin glasses \cite{knuth74}.

For $J^{(N,d)}$, the parameter $d$ significantly influences the landscape's ruggedness. As $d$ decreases (e.g., $d = -0.5, 0.2$), the number of local minima increases substantially. For instance, at $N=28$, instances with $d=-0.5$ possess over $10^7$ local minima, whereas instances with $d=2$ have approximately $3 \times 10^5$ local minima. This tunability of landscape complexity via the parameter $d$ is a key feature of our proposed $J^{(N,d)}$.

Comparing our results to the Wishart planted ensemble (dashed lines in Figure~\ref{fig:local_minima}), we observe that our deterministic instances, particularly for $d \le 0.5$, can exhibit a number of local minima that is comparable to, or even an order of magnitude greater than of those found in the Wishart ensemble for typical $\alpha$ values at the same system size. For example, $J^{(N=20,~d=-0.5)}$ shows approximately $5 \times 10^5$ local minima, while the Wishart ensemble with $\alpha=0.1$ (a parameter regime known for its difficulty in that model) exhibits around $2 \times 10^4$ local minima. This suggests that the energy landscapes generated by $J^{(N,d)}$ are indeed highly complex and, by this measure of local optima count, can be even more rugged than these established random benchmarks.

\subsection{Full Energy Spectrum and Density of States}
\label{subsec:full_energy_spectrum}

\begin{figure*}[htbp!]
    \centering
    \begin{subfigure}[b]{0.38\textwidth}
        \centering
        \includegraphics[width=\textwidth]{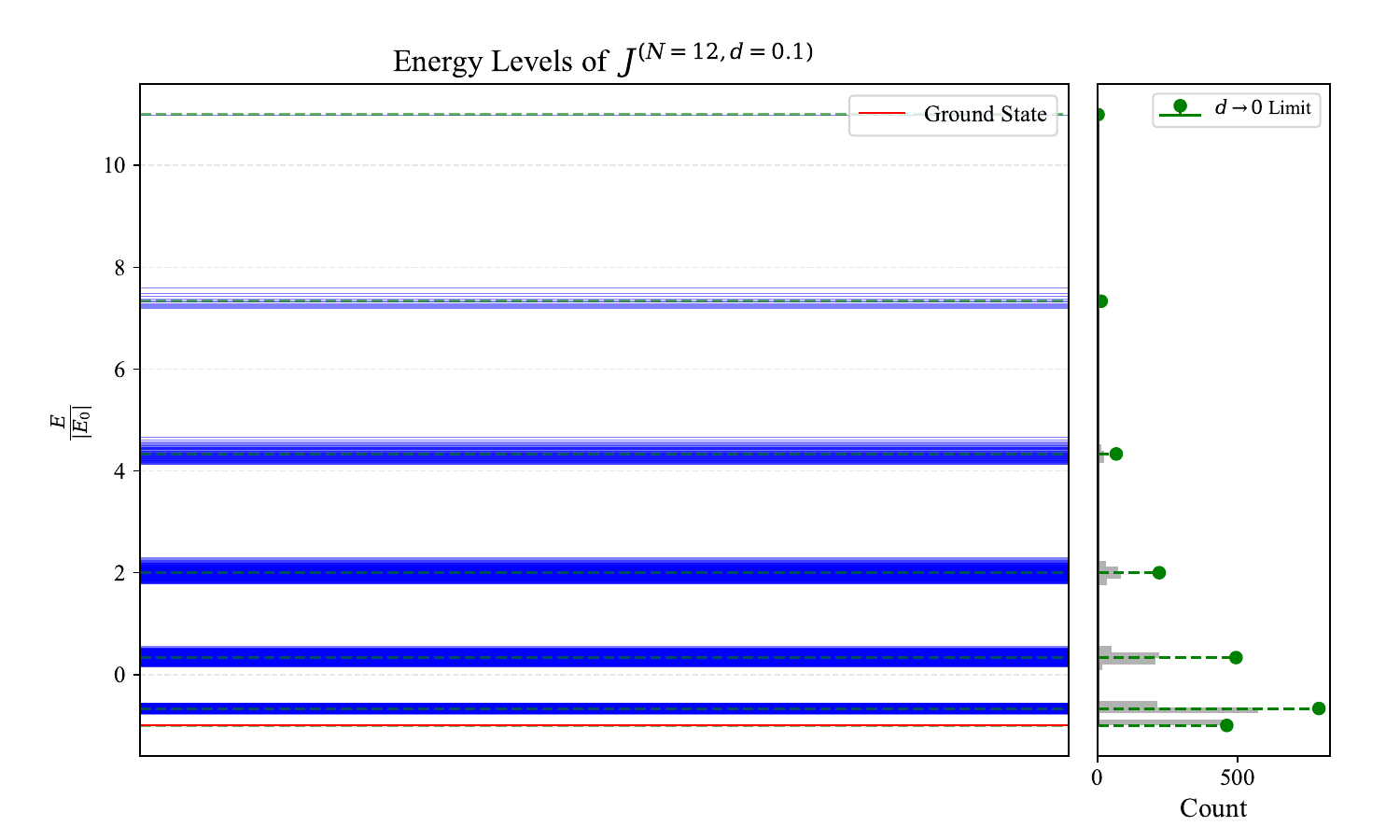} % REUSE FILENAME, content interpretation changes
        \caption{$d=0.1$}
        \label{fig:subfig_J_d01_fullspec}
    \end{subfigure}
    % \hfill %
    \begin{subfigure}[b]{0.38\textwidth}
        \centering
        \includegraphics[width=\textwidth]{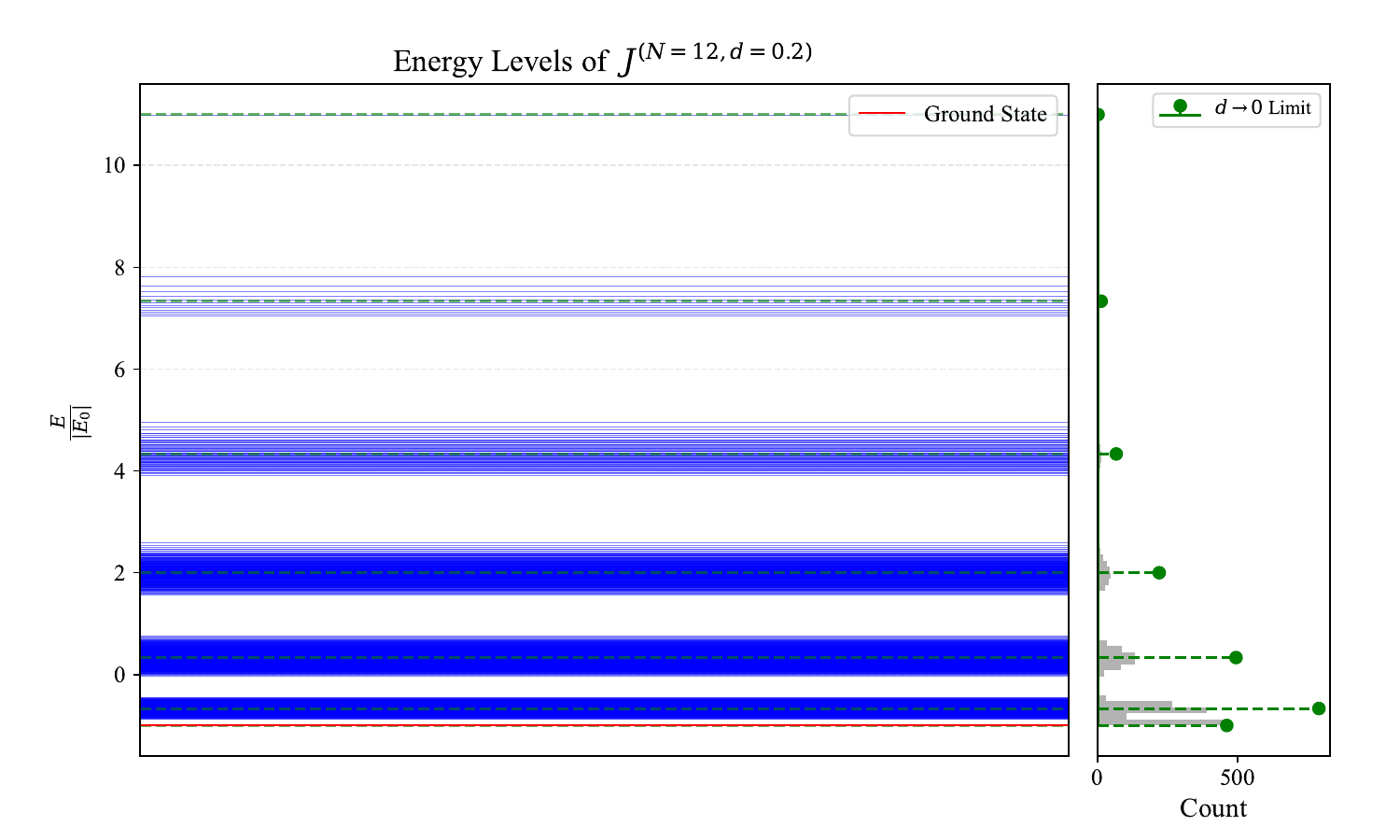} % REUSE FILENAME
        \caption{$d=0.2$}
        \label{fig:subfig_J_d02_fullspec}
    \end{subfigure}
    % \vspace{0.5cm} % Optional vertical spacing
    \begin{subfigure}[b]{0.38\textwidth}
        \centering
        \includegraphics[width=\textwidth]{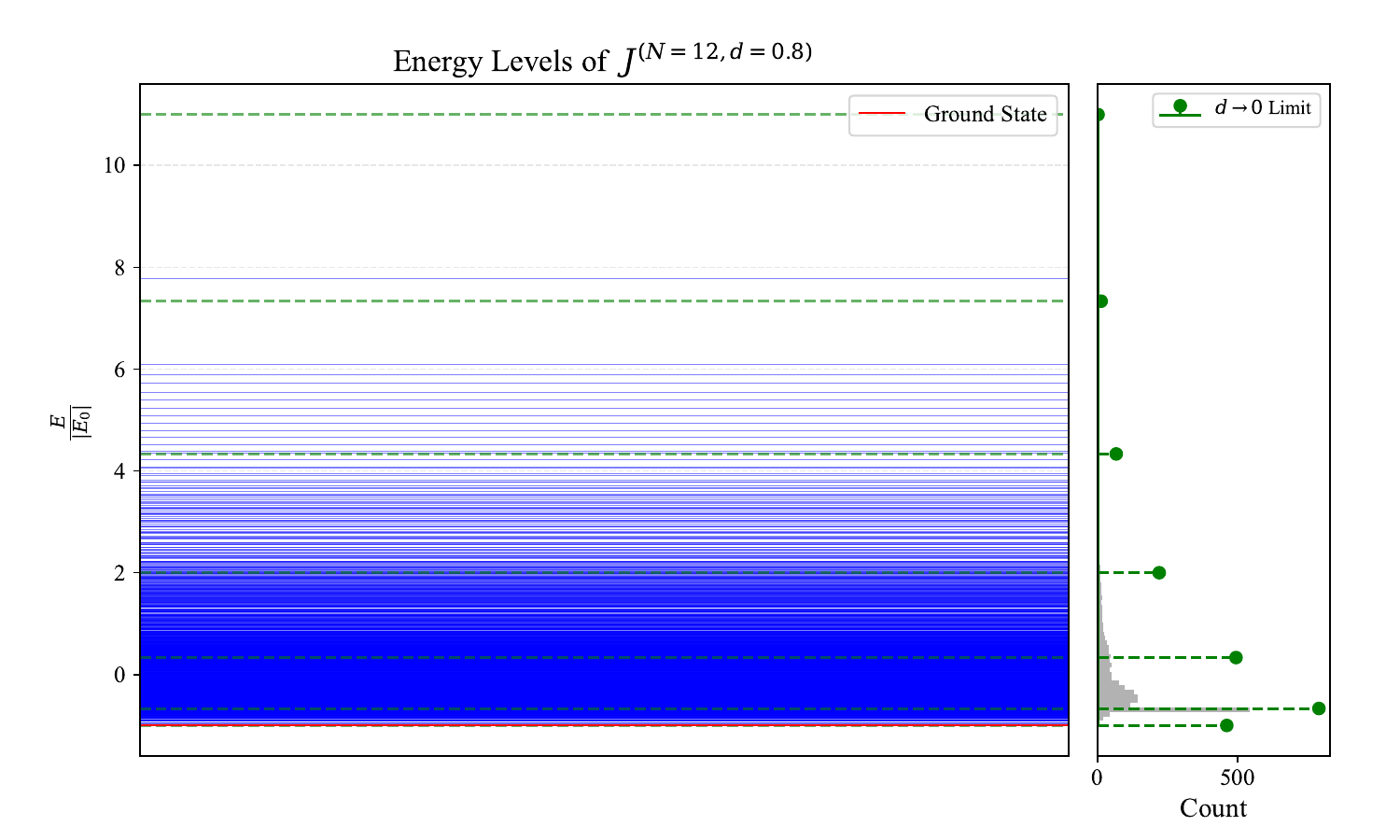} % REUSE FILENAME
        \caption{$d=0.8$}
        \label{fig:subfig_J_d08_fullspec}
    \end{subfigure}
    % \hfill %
    \begin{subfigure}[b]{0.38\textwidth}
        \centering
        \includegraphics[width=\textwidth]{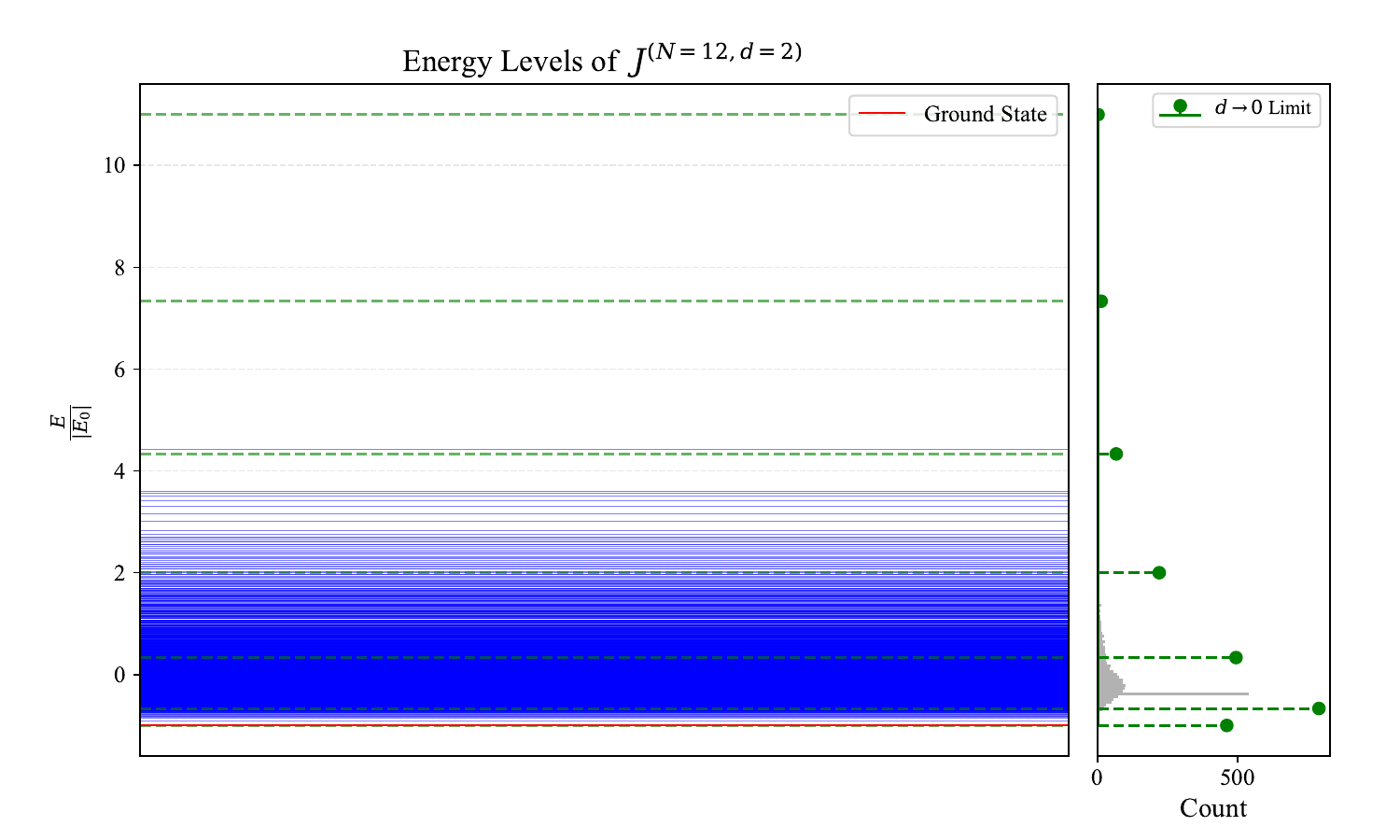} % REUSE FILENAME
        \caption{$d=2$}
        \label{fig:subfig_J_d20_fullspec}
    \end{subfigure}
    \caption{Full energy spectrum and density of states (DOS) for our proposed model $J^{(N,d)}$ with $N=12$ and varying $d$. The main panel shows all distinct energy levels (blue lines, normalized by $|E_0(N,d)|$) with the ground state in red. Green dashed lines represent the "$d \to 0$ Limit" (theoretical levels for uniform coupling, see text and Eq.~\ref{eq:degen_k_d0}). The right sub-panel in each plot is a histogram representing the DOS (count of configurations per energy bin).}
    \label{fig:full_spectrum_J_nd_panel}
\end{figure*}

\begin{figure*}[htbp!]
    \centering
    \begin{subfigure}[b]{0.38\textwidth}
        \centering
        \includegraphics[width=\textwidth]{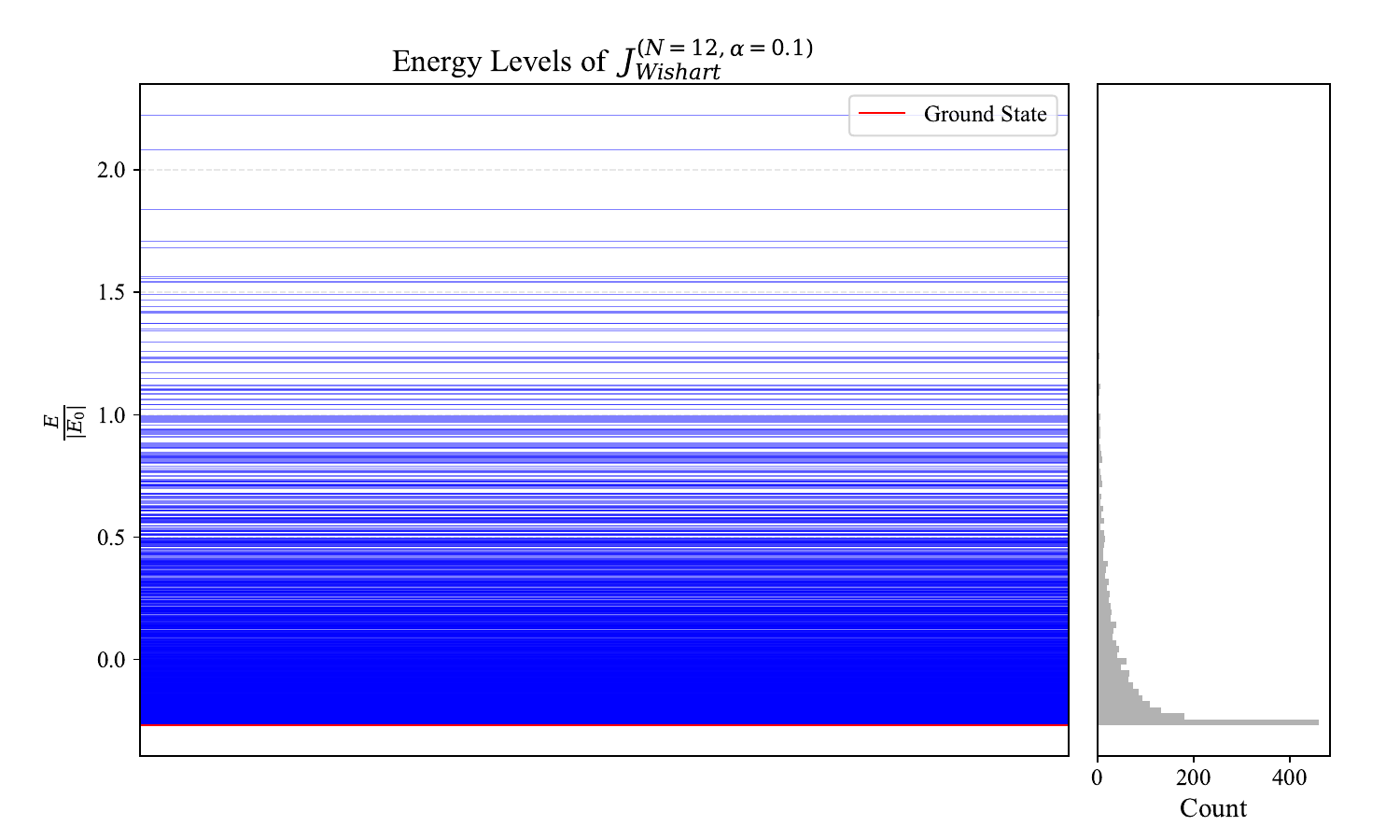} % REUSE FILENAME
        \caption{$\alpha=0.1$}
        \label{fig:subfig_wishart_alpha01_fullspec}
    \end{subfigure}
    % \hfill %
    \begin{subfigure}[b]{0.38\textwidth}
        \centering
        \includegraphics[width=\textwidth]{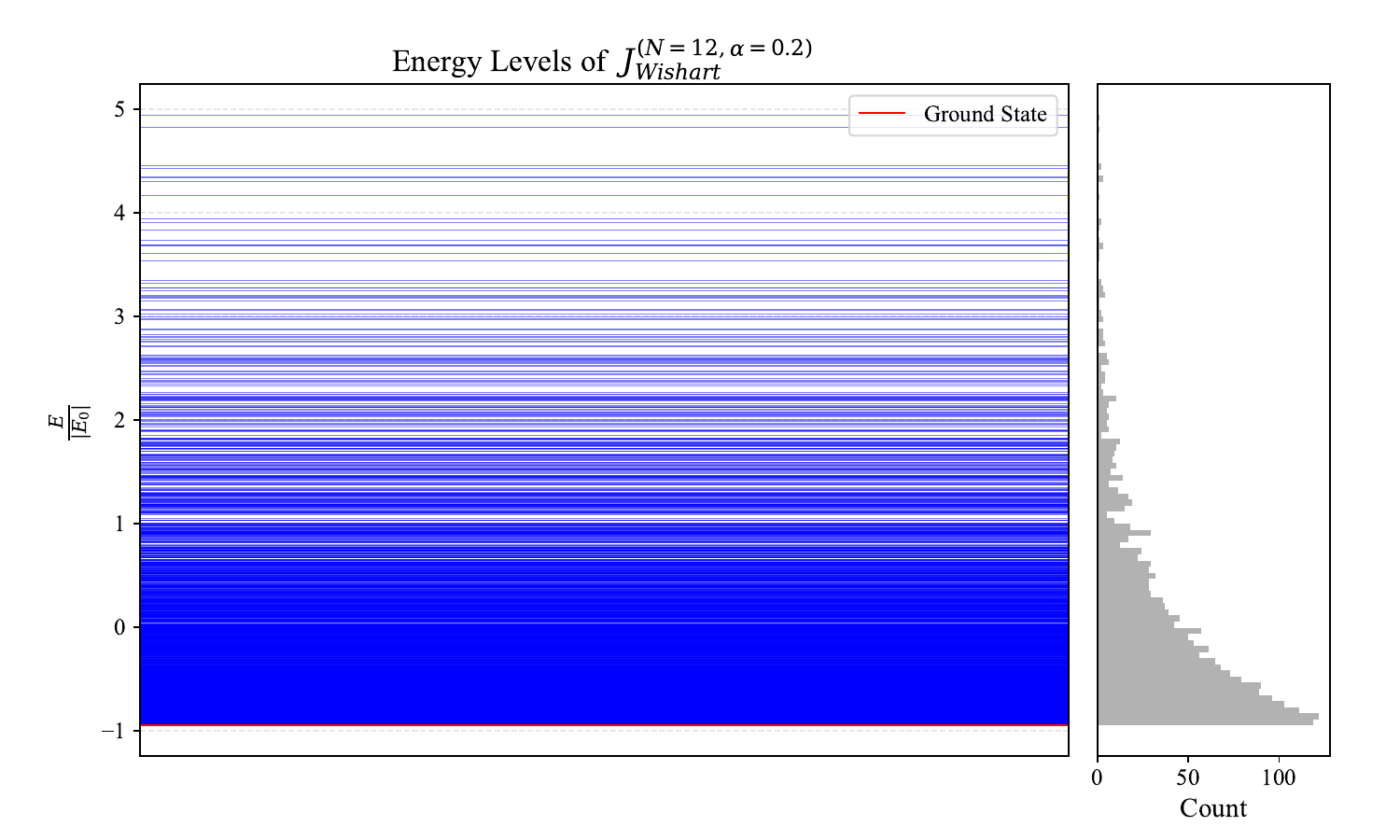} % REUSE FILENAME
        \caption{$\alpha=0.2$}
        \label{fig:subfig_wishart_alpha02_fullspec}
    \end{subfigure}
    % \vspace{0.5cm} % Optional vertical spacing
    \begin{subfigure}[b]{0.38\textwidth}
        \centering
        \includegraphics[width=\textwidth]{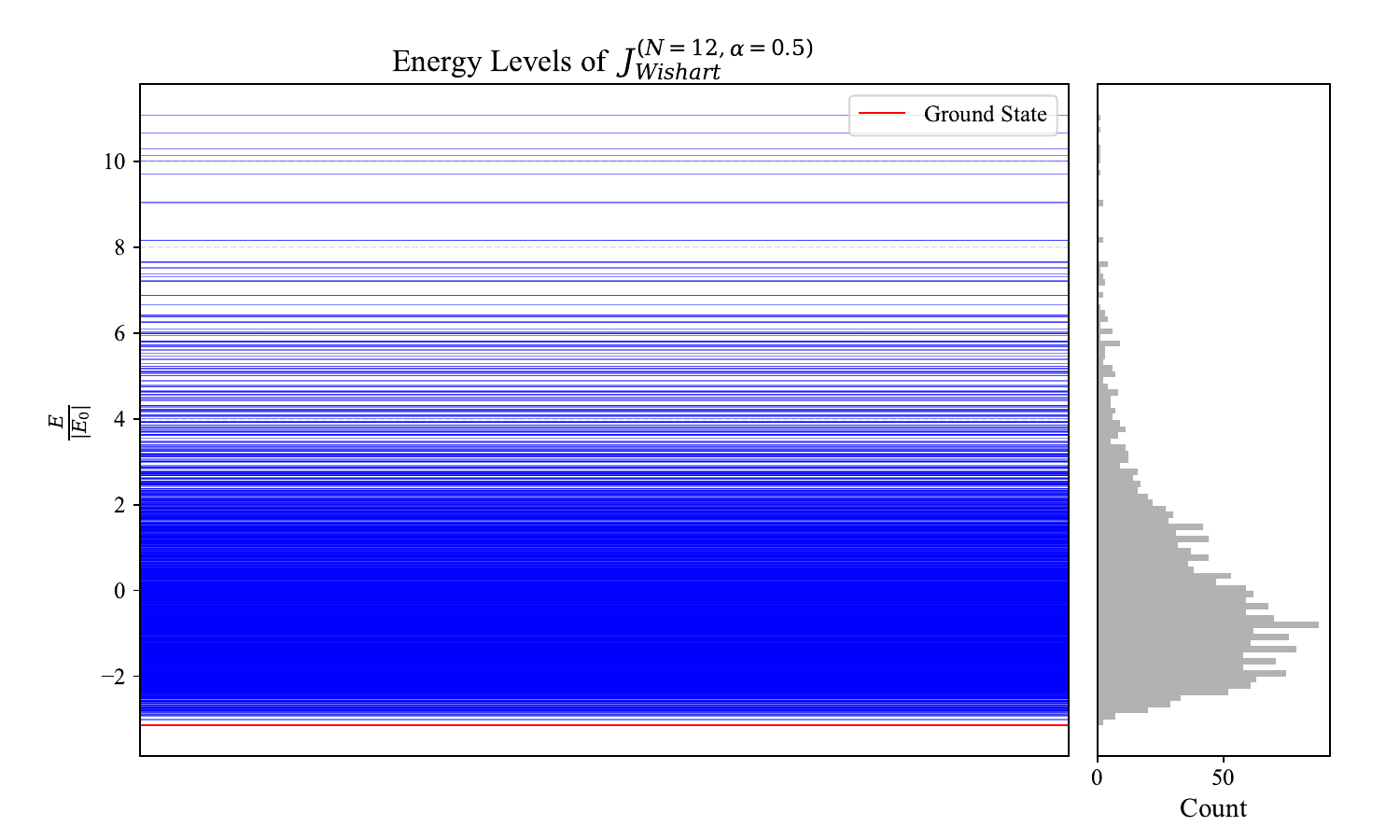} % REUSE FILENAME
        \caption{$\alpha=0.5$}
        \label{fig:subfig_wishart_alpha05_fullspec}
    \end{subfigure}
    % \hfill %
    \begin{subfigure}[b]{0.38\textwidth}
        \centering
        \includegraphics[width=\textwidth]{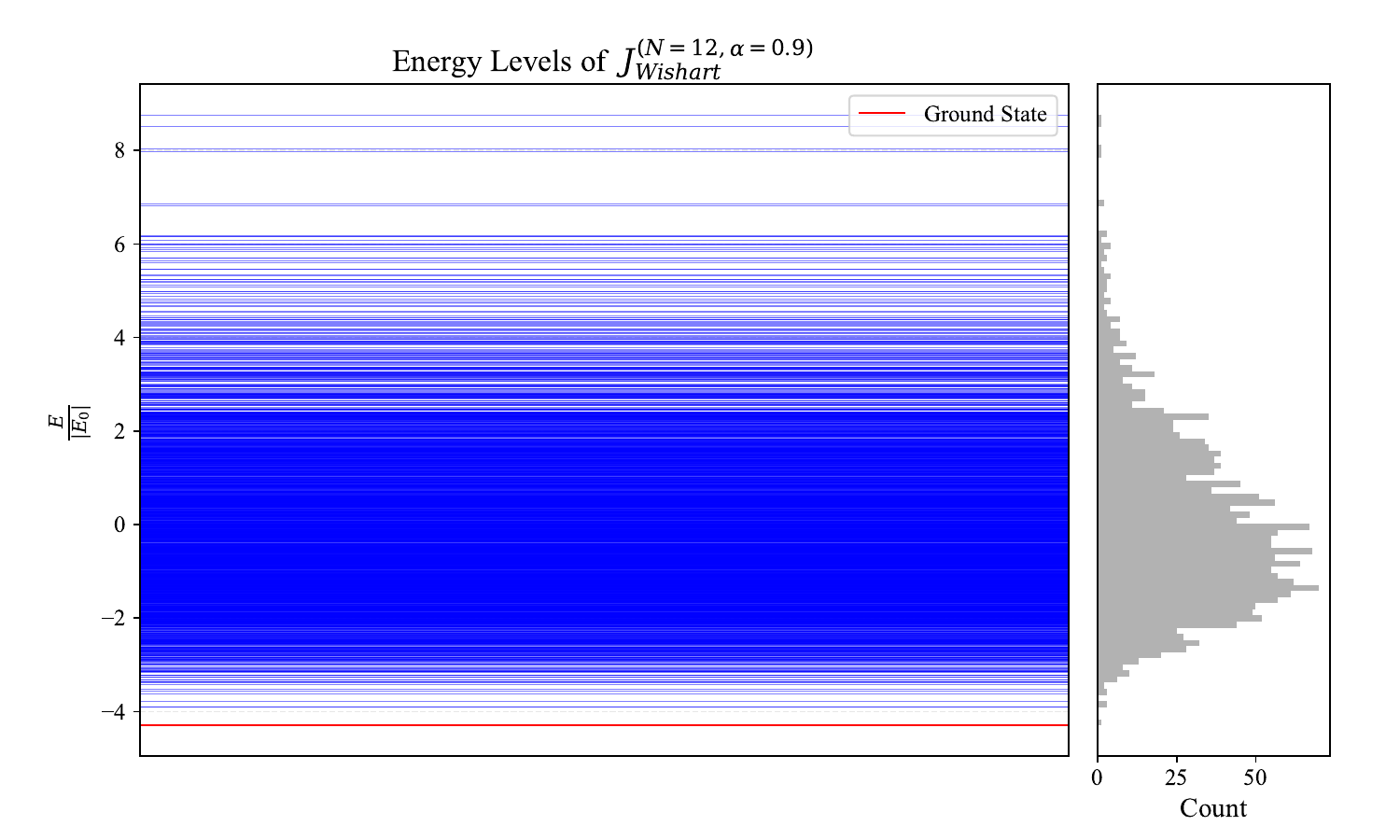} % REUSE FILENAME
        \caption{$\alpha=0.9$}
        \label{fig:subfig_wishart_alpha09_fullspec}
    \end{subfigure}
    \caption{Full energy spectrum and density of states (DOS) for the Wishart planted ensemble $J_{Wishart}$ with $N=12$ and varying $\alpha$. The main panel shows all distinct energy levels (blue lines, normalized by $|E_0(N,\alpha)|$) with the ground state in red. The right sub-panel in each plot is a histogram representing the DOS (count of configurations per energy bin).}
    \label{fig:full_spectrum_Wishart_panel}
\end{figure*}

To further characterize the energy landscape beyond identifying local optima, we examine the full energy spectrum, encompassing all $2^N$ possible spin configurations, for $J^{(N,d)}$ and for the Wishart ensemble. Understanding the distribution of all energy levels provides insights into the overall structure of the landscape and the statistical properties of typical states. For the visualizations presented (Figures~\ref{fig:full_spectrum_J_nd_panel} and \ref{fig:full_spectrum_Wishart_panel}), energies $E$ of all configurations for a given instance are normalized by the magnitude of its specific ground state energy, $|E_0|$. This allows for a consistent comparison of the energy scales relative to the depth of the global minimum for each case. Each main panel displays all distinct energy levels found among the $2^N$ states, while the accompanying sub-panel on the right provides a histogram illustrating the density of states (DOS) across the energy spectrum (i.e., the count of configurations per energy bin).

Figure~\ref{fig:full_spectrum_J_nd_panel} displays these full energy spectra and their corresponding DOS for $J^{(N,d)}$ with $N=12$ and varying $d$.
A striking feature, particularly when $d$ is small (Figures~\ref{fig:full_spectrum_J_nd_panel}a for $d=0.1$ and \ref{fig:full_spectrum_J_nd_panel}b for $d=0.2$), is the appearance of highly degenerate energy levels, forming distinct bands in the spectrum. These bands are prominent in both the level display and as sharp peaks in the DOS.

The green dashed lines in these plots, labeled "$d \to 0$ Limit," represent the positions of the highly degenerate macroscopic energy levels characteristic of a fully-connected Ising model with uniform ferromagnetic couplings (e.g., $J_{ij} = C_0(1-\delta_{ij})$ for some constant $C_0>0$). This corresponds to the $d=0$ limit of our interaction. In such a model, the energy $E_k^{(0)}$ of a state depends only on its total magnetization $M_{k} = N - 2k$ (where $k=0,1,\dots,\lfloor N/2\rfloor$). The number of distinct spin configurations (degeneracy) $g_k$ for each such energy level $E_k^{(0)}$ is given by the well-known combinatorial factor:
\begin{equation}
    g_k = \begin{cases}
            2 \binom{N}{k}, & \text{if } M_{k} \neq 0, \\
            \binom{N}{N/2}, & \text{if } M_{k} = 0 \text{ (N even)}.
          \end{cases} \label{eq:degen_k_d0}
\end{equation}
The factor of 2 for $M_{k} \neq 0$ accounts for spin-flip symmetry. The positions of the green "$d \to 0$ Limit" lines in Figure~\ref{fig:full_spectrum_J_nd_panel} are determined by the energies $E_k^{(0)}$ (normalized by $|E_0(N, d=0)|$).
Remarkably, for small $d$, the density of states (DOS) for $J^{(N,d)}$, shown by the histograms, exhibits pronounced peaks that align closely with these "$d \to 0$ Limit" energy levels. This alignment indicates that even with non-uniform couplings (for small $d$), $J^{(N,d)}$ retains a structure where large numbers of states cluster around energy values characteristic of the highly degenerate constant-coupling model. The heights of the peaks in our DOS reflect the high count of states within these energy bands, echoing the large degeneracies $g_k$ of the $d=0$ limit.

As $d$ increases (Figures~\ref{fig:full_spectrum_J_nd_panel}c for $d=0.8$ and \ref{fig:full_spectrum_J_nd_panel}d for $d=2$), the influence of the non-uniform $(i^d+j^d)$ terms becomes more pronounced. The distinct energy bands broaden, and the gaps between them become less defined, leading to a more quasi-continuous DOS. While the "$d \to 0$ Limit" lines are still shown for reference, their direct correspondence to sharp peaks in the DOS diminishes. However, the overall DOS still shows a significant concentration of states at energies above the ground state. To better visualize the transition, an animation showing the continuous evolution of the energy spectrum and DOS as $d$ varies from negative to positive values is available in the Supplemental Material \cite{supplementary}.

Figure~\ref{fig:full_spectrum_Wishart_panel} shows the corresponding full energy spectra and DOS for the Wishart planted ensemble with $N=12$ and varying $\alpha$. In contrast to $J^{(N,d)}$ with small $d$, the Wishart instances generally exhibit a much more continuous-looking spectrum of energy levels across all $\alpha$ values shown (Figures~\ref{fig:full_spectrum_Wishart_panel}a-d). The DOS typically forms a broader, more Gaussian-like distribution (especially for larger $\alpha$, e.g., $\alpha=0.9$, Figure~\ref{fig:full_spectrum_Wishart_panel}d), without the sharp, discrete peaks associated with the highly degenerate bands seen in $J^{(N,d)}$ for small $d$. For small $\alpha$ (e.g., $\alpha=0.1$, Figure~\ref{fig:full_spectrum_Wishart_panel}a), the DOS is more skewed, with a higher density of states closer to the ground state energy, but still lacks the distinct banding.

This analysis of the full energy spectrum reveals fundamental structural differences in the energy landscapes. $J^{(N,d)}$, particularly for small $d$, exhibits a landscape characterized by highly degenerate energy shells, a feature directly inherited from the $d \to 0$ limit of uniform coupling, as described by Equation (\ref{eq:degen_k_d0}). This results in a "staircase-like" DOS. The Wishart ensemble, being derived from random matrices, generally presents a smoother, more continuous DOS. These differences in the overall distribution of available energy states likely have profound implications for the thermodynamics and search dynamics within these landscapes, distinguishing our deterministic construction from typical random ensembles. The high degeneracy within specific energy bands in $J^{(N,d)}$ might offer unique pathways or challenges for annealing algorithms compared to landscapes with a more continuous DOS.

\section{Conclusion}
In this study, we introduced a novel class of fully connected Ising models. To analytically solve this class, we reformulated the discrete Ising Hamiltonian into a continuous framework, enabling us to determine the exact ground state. This reformulation represents a significant advancement in the analytical treatment of complex Ising systems.

Our analytical solutions were validated through numerical experiments with brute-force calculations, the Simulated Coherent Ising Machine (SimCIM), and the D-Wave quantum computer. The results demonstrated perfect agreement between brute-force calculations and our method for small-scale systems, as well as between SimCIM and our approach for larger systems. However, significant deviations were observed in the D-Wave quantum solver’s results, as shown in Figure \ref{fig:dwave}. These deviations were initially examined for small-scale problems, and the analysis was extended to larger problem sizes, where the deviations persisted and became more pronounced, highlighting limitations in current quantum hardware for this class of problems.

The continuous formulation of the Ising Hamiltonian presented herein not only offers a pathway for analytically solving a new class of Ising problems but also deepens our understanding of how inherent order influences collective behavior. The emergence of a predictable, two-cluster ground state for our defined $J_{ij}^{(N,d)}$ can be intuitively grasped by recalling the notion of a system with graded component properties---akin to our 'city' model where 'economic power' dictates interaction strengths. While the energy landscape of $J^{(N,d)}$ can still exhibit a rich structure with numerous local minima (as discussed in Sec.~\ref{sec:energy_landscape}), the deterministic nature of its couplings, tied to an inherent ordering, allows for an analytical prediction of its global ground state. This presents a different character of complexity compared to fully disordered systems where identifying the ground state is often computationally prohibitive. Beyond its utility in advancing analytical techniques for quantum simulation and computation, this work underscores the distinct physics arising from such ordered interactions and provides a robust, analytically grounded testbed for assessing the fidelity of Ising minimizers, thereby circumventing the need for computationally expensive brute-force validation for this class of problems.

\section*{Author Contributions}
Amirhossein Rezaei developed the main theoretical ideas, mathematical proofs, and the overall framework of the study. Mahmood Hasani and Alireza Rezaei contributed to the writing, formatting, and refinement of the manuscript, and performed benchmarking of Ising Machines. S.M. Hassan Halataei provided institutional support.

\section*{Acknowledgment}
We would like to express our gratitude to Dr. Behrouz Askari for his valuable comments and to Dr. G. Reza Jafari for his insightful contributions during the revision process. Their input was appreciated and contributed to the development of this paper.

\bibliography{apssamp.bib}% Produces the bibliography via BibTeX.

\appendix
\section{CIM Hyperparameters}\label{appendixa}
The hyperparameters for SimCIM were tuned with Bayesian optimization. The hyperspace for finding the ground state is shown in Table \ref{table:1}. In this table, the hyperparameters CAC-$\alpha$ and CAC-$\beta$ were optimized within a range of $\pm20\%$ of their initial values, while CAC-$\tau$ and CAC-$\gamma$ were manually set and were not subject to optimization. The time span was fixed at 10000 to ensure we reach the most optimal solution. This approach allowed us to reach the ground state for different values of $d$.

\begin{table}[H]
\centering
\begin{tabular}{||c c c||} 
 \hline
 Hyperparameter & Value & Search Space\\ [0.5ex] 
 \hline\hline
 CAC-$\alpha$ & 0.7 & [0.56, 0.84] ($\pm20\%$)\\ 
 CAC-$\beta$ & 0.25 & [0.2, 0.3] ($\pm20\%$)\\
 CAC-$\tau$ & 150 & Manually set\\
 CAC-$\gamma$ & 0.01 & Manually set\\[1ex] 
 \hline
\end{tabular}
\caption{Hyperspace of the Simulated Coherent Ising Machine with Bayesian Optimization}
\label{table:1}
\end{table}

\section{Evaluating Integral}\label{Evaluating Integral}
By expanding Equation (\ref{eq:hintegral}), we can obtain
\begin{multline} \label{eq:double}
    H_{\Lambda}(d, \mathbf{q}) =  \int_{0}^{1}\int_{0}^{1} \frac{N^2}{2} x^d \prod_{{\alpha}=1}^{\Lambda}\sgn(x-q_{\alpha})\sgn(y-q_{\alpha}) \; dx \; dy + \\ \int_{0}^{1}\int_{0}^{1} \frac{N^2}{2} y^d \prod_{{\alpha}=1}^{\Lambda}\sgn(x-q_{\alpha})\sgn(y-q_{\alpha}) \; dx \; dy.
\end{multline}

Both parts are the same integral, except that the symbols $x$
 and $y$
 are swapped. We can therefore rewrite Equation (\ref{eq:double}) as:
 \begin{equation} \label{eq:same}
     H_{\Lambda}(d, \mathbf{q}) =  N^2 \int_{0}^{1}\int_{0}^{1} x^d \prod_{\alpha=1}^{\Lambda}\sgn(x-q_\alpha)\sgn(y-q_\alpha) \; dx \; dy.
 \end{equation}
 
Using Fubini's Theorem, we obtain:
\begin{equation}
    H_{\Lambda}(d, \mathbf{q}) = N^2 \int_{0}^{1} x^d \prod_{\alpha=1}^{\Lambda}\sgn(x-q_\alpha) \; dx \int_{0}^{1} \sgn(y-q_\alpha) \; dy.
\end{equation}

Now we can omit the sign functions and by considering $q_0$ and $q_{\Lambda+1}$, we have:
\begin{equation}
    H_{\Lambda}(d, \mathbf{q}) = N^2 \left(\sum_{\alpha=0}^{\Lambda} \int_{q_{\alpha}}^{q_{\alpha+1}} x^d(-1)^{\Lambda - \alpha} \; dx\right) \left(\sum_{\alpha=0}^{\Lambda} \int_{q_\alpha}^{q_{\alpha+1}}(-1)^{\Lambda-\alpha}\; dy\right),
\end{equation}
where this integral could be solved as bellow:
\begin{equation}
    H_{\Lambda}(d, \mathbf{q}) = N^2 \left(\sum_{\alpha=0}^\Lambda\left[\frac{x^{d+1}}{d+1}(-1)^{\Lambda-\alpha}\right]_{q_\alpha}^{q_{\alpha+1}}\right)\left(\sum_{\alpha=0}^\Lambda\left[y(-1)^{\Lambda-\alpha}\right]_{q_\alpha}^{q_{\alpha+1}}\right),
\end{equation}
\noindent which leads to Equation (\ref{eq:finalhamil}) $\square$.
\section{Uniqueness of Ground State Pattern} \label{app:unique}
By taking the derivative of $H_1$ with respect to $q_1$ and equating it to $0$ we obtain:
\begin{equation} \label{eq:h1derive}
    2q_1^{d+1} - 1 = - (d+1)q_1^d(2q_1-1).
\end{equation}

Using Equation (\ref{eq:h1derive}) and Equation (\ref{eq:finalhamil}), we can write:
\begin{equation}
    H_1(d,q_1) = -N^2(2q_1-1)^2q_1^d,
\end{equation}
which is always a negative number. For $d>0$, it can be shown that for each $i$, we have $q_i > \frac{1}{2}$. From Equation (\ref{eq:h1derive}), we can write:
\begin{equation} \label{eq:h1d}
    2q_1^{d+1} + (d+1)q_1^d(2q_1-1) = 1. \\
\end{equation}

By reordering the left hand side of Equation (\ref{eq:h1d}) :
\begin{equation} \label{eq:h1derivorder}
    q_1^d(2q_1 + (d+1)(2q_1-1)) = 1.
\end{equation}
The expression in parenthesis on the left hand side of Equation (\ref{eq:h1derivorder}) is strictly greater than 1 for $d>0$:
\begin{equation}
    2q_1 + (d+1)(2q_1-1) > 1,
\end{equation}
which can be simplified further:
\begin{equation}
    (2q_1-1)(d+2) > 0,
\end{equation}
from which we can readily verify that $q_1 > \frac{1}{2}$ for $d>0$. 
Using this result, it can be shown that $H_1(d,q_1)$ is convex on the interval $q_1 > \frac{1}{2}$ and $d>0$, by taking its second derivative:
\begin{equation} \label{eq:seconderiv}
    \frac{\partial^2 H_1}{\partial q_1^2} = 2 q_1^{d-1}(4q_1 + d(2q_1 -1)),
\end{equation}
and observing that for $q_1 > \frac{1}{2}$ and $d>0$, Equation (\ref{eq:seconderiv}) is strictly positive. The convexity of $H_1(d,q_1)$ on interval $q_1 > \frac{1}{2}$ and $d>0$, implies the uniqueness of $q_1$. This can be shown by observing that $\frac{\partial H_1}{\partial q_1}|_{q_1=\frac{1}{2}} < 0$ and $\frac{\partial H_1}{\partial q_1}|_{q_1=1} > 0$. Thus, the mean value theorem implies that there exists a zero for the first derivative of $H_1(d,q_1)$ with respect to $q_1$, on interval $[\frac{1}{2},1]$ and by convexity of $H_1(d,q_1)$ on the interval, we can conclude that the root is unique.

\section{Finding Ground State Pattern} \label{app:pattern}
For $\Lambda \ge 2$, first we find the critical points of Equation (\ref{eq:finalhamil}). Taking the derivative, with respect to $q_j$ results in Equation (\ref{eq:deriv}):
\begin{multline} \label{eq:deriv}
    \frac{\partial H_{\Lambda}}{\partial q_j} = (-1)^{j+1} 2 q_j^d \left((-1)^\Lambda + 2\sum_{i=1}^\Lambda (-1)^{i+1}q_i\right)+ \\(-1)^{j+1} \frac{2}{1+d} \left((-1)^\Lambda + 2\sum_{i=1}^\Lambda (-1)^{i+1}q_i^{d+1}\right).
\end{multline}

At a critical point all of the derivatives, as shown in Equation (\ref{eq:deriv}), must be zero. Consequently, for $\Lambda\ge 2$ and $j \in [1,\Lambda-1]$, sum of the derivatives of Equation (\ref{eq:finalhamil}) with respect to $q_j$ and $q_{j+1}$ must also be 0. Thus, we can write:
\begin{multline} \label{eq:sumderiv}
    \frac{\partial H_{\Lambda}}{\partial q_j}  + \frac{\partial H_{\Lambda}}{\partial q_{j+1}} = 2(-1)^j\left(q_j^d-q_{j+1}^d\right)\times \\
    \left[(-1)^\Lambda + 2\sum_{n = 1}^\Lambda(-1)^{n+1}q_n\right] = 0.
\end{multline}

For Equation (\ref{eq:sumderiv}) to hold, we must have:
\begin{equation} \label{eq:condition}
    (-1)^\Lambda + 2\sum_{n=1}^\Lambda(-1)^{n+1} q_n = 0.
\end{equation}

Equation (\ref{eq:condition}) implies that, the Hamiltonian in Equation (\ref{eq:finalhamil}) in critical points for $\Lambda\ge 2$, is 0. The only remaining interesting points are the boundaries of the domain of the Hamiltonian. The first two boundaries is to set $q_1 = 0$ and $q_\Lambda = 1$. First we show that if we have $q_\Lambda \rightarrow 1$, then $H_{\Lambda}(q_\Lambda \rightarrow 1) \rightarrow H_{\Lambda-1}$:

\begin{multline}
        H_{\Lambda}(d, q_\Lambda \rightarrow 1) = \frac{N^2}{1 + d}\left((-1)^\Lambda + 2(-1)^{\Lambda+1} + 2\sum_{i=1}^{\Lambda-1} (-1)^{i+1}q_i\right)\\ \left((-1)^\Lambda + 2(-1)^{\Lambda+1} +2\sum_{i=1}^{\Lambda-1} (-1)^{i+1}q_i^{d+1}\right),
\end{multline}
and this can be simplified as:
\begin{multline}
        H_{\Lambda}(d, q_\Lambda \rightarrow 1) = \frac{N^2}{1 + d}\left((-1)^{\Lambda-1} + 2\sum_{i=1}^{\Lambda-1} (-1)^{i+1}q_i\right)\times \\ \left((-1)^{\Lambda-1} +2\sum_{i=1}^{\Lambda-1} (-1)^{i+1}q_i^{d+1}\right),
\end{multline}
which is equal to $H_{\Lambda-1}(d,\mathbf{q})$. Since the same argument for the critical points of $H_{\Lambda}$ can also be made for $H_{\Lambda-1}$, we can conclude that the global minimum of $H_{\Lambda-1}$ must also lie on its boundary and the same argument can be made for $H_{\Lambda-2}$ by letting $q_{\Lambda-1} \rightarrow 1$ and obtaining $H_{\Lambda-2}$. We may continue in this manner, until we reach $H_1$, which its minimum value is given in Equation (\ref{eq:12}). The same procedure can be shown when $q_1 \rightarrow 0$, in which case, we also have $H_{\Lambda} \rightarrow H_{\Lambda-1}$, with the difference that the indices for $q_i$s shift by one, i.e., in the new Hamiltonian, $q_{i+1} \rightarrow q_i$.

For $d \le -1$, Equation (\ref{eq:hintegral}) does not converge, as it has a singularity at $x = y = 0$. To overcome this issue, we can simply keep the lower limit of integral in Eq. (\ref{eq:hintegral}) as $\frac{1}{N}$ and avoid the singularity. Without any loss of generality, and by a similar approach as before, we can find the extended expression for $q(d)$:
\begin{equation}\label{extended_q}
(1+(\frac{1}{N})^{d+1}-2 q^{d+1})+(d+1) (1+ \frac{1}{N} -2 q) q^d = 0.
\end{equation}

The proof for this case follows a similar line of reasoning as the proof for $d > -1$, with appropriate adjustments for the modified lower limit of the integral. While this extended form is more general, we chose to present the proof for $d > -1$ in detail as it is more concise and illustrates the key principles without the additional complexity introduced by the regularization term. 
\section{Permutation Invariance of $J^{(N, d)}$}
Here, we examine another property of this class of interaction matrices. This characteristic allows us to determine the interaction matrix for any given spin configuration, assuming that the spin configuration represents the ground state of the interaction matrix. This is achieved by exploiting the properties of $J^{N, d}$. Rewriting Equation (\ref{eq:genhamil}) in vector-matrix notation (and discarding the factor of $\frac{1}{2}$), we have:
\begin{equation}
    H = \pmb{s}^T J \pmb{s}.
\end{equation}

Now note that $H$ is invariant under the following transformation:

\begin{equation}\label{eq14} 
    H = \pmb{s}^T P^{T} P J P^{T} P \pmb{s} = \pmb{s}'^T J' \pmb{s}',
\end{equation}
where $P$ is a permutation matrix. As we have shown, the ratio $q$ can be any value between $[0, 1]$. Considering this, the $\mathbb{Z}_2$ symmetry and the permutation invariance of $J$, \textit{any} configuration of $\pmb{s}$ can be represented as the ground state of $J^{N, d}$. To further illustrate this, note that any configuration $\pmb{s}$, has fixed number of up and down spins, and the ratio of spins to the system size ($q$) is always between 0 and 1. Assume that the said configuration has two clusters of up and down spins, and is not scrambled. Now to find the proper interaction matrix for the ratio $q$, we can simply select $d$ in accordance with Equation (\ref{extended_q}) and Figure \ref{fig2}. Now, to unscramble the configuration, we can use permutation matrices repeatedly. This allows us to sort this configuration to two clusters of up and down spins (using Equation (\ref{eq14})) and with that, we can also sort the interaction matrix.

\end{document}